\documentclass[letterpaper,preprintnumbers,twocolumn,superscriptaddress,aps,nofootinbib,perprint]{revtex4}
\usepackage{CJK}\usepackage{amssymb}\usepackage[centertags]{amsmath}\usepackage{txfonts}\usepackage{epsfig}\usepackage{bm}\usepackage{color}\usepackage{graphicx,graphics}\usepackage{multirow}\usepackage{float}\usepackage{ulem}\usepackage{hyperref}\usepackage{setspace}\usepackage{slashed}
\usepackage{booktabs}\usepackage{geometry}\usepackage{diagbox}\usepackage{epstopdf}
\hypersetup{colorlinks=true, citecolor=blue, linkcolor=blue,filecolor=black,urlcolor=blue}
\allowdisplaybreaks[4]

\begin{document}

\title{The forward-backward asymmetry in the electron positron annihilation process at twist-4}

\author{Weihua Yang}

\affiliation{College of Nuclear Equipment and Nuclear Engineering, Yan Tai University, Yantai, Shandong 264005, China}

\author{Chao Li}

\affiliation{School of Mathematics and Physics, Nanyang Institute of Technology, Nanyang, 473000, China}

\begin{abstract}
A quantity of particular experimental interest is the forward-backward asymmetry in the angular distribution of positively and negatively charged fermions produced in $Z^0$ decays.  Measurements of this asymmetry can enable independent determinations of the neutral-current couplings of these fermions, i.e. the $Z^0$ boson couplings for left- and right-handed fermions, respectively.  Due to the quark confinement, however,  it is difficult to determine the electroweak interactions of quarks, especially for light quarks. In the hadron production electron positron annihilation process, the parton model with factorization theorem gives a reliable approximate description. Quantities are thus expressed in terms of fragmentation functions in the annihilation process.  In this paper, we consider the vector meson production in the inclusive electron positron annihilation process and calculate the forward-backward asymmetry in the hadronic level. Calculations are carried out by applying the collinear expansion in the parton model at leading order twist-4.  We note here this process provides not only  a tool for analyzing the hadronic weak interactions but also an opportunity for understanding the parton model of the strong interaction. In other words, the results can be used to test the electroweak and strong interactions simultaneously.  

\end{abstract}

\maketitle

\section{Introduction}

The Standard Model (SM) of elementary particles and their interactions has achieved great success in the past decades. It has two basic components, the spontaneously broken electroweak (EW) theory and the color gauge theory or quantum chromodynamics (QCD).  Since the left- and right-handed fermions in the EW theory live in different representations of the fundamental gauge group, they have different couplings for the gauge bosons, $Z^0, W^\pm$. As for $Z^0$, the difference leads to an asymmetry in the angular distribution of positively and negatively charged leptons and/or quarks produced in Z boson decays. This asymmetry, known as the forward-backward asymmetry~\cite{Wu:1984ik,Marshall:1988bk,Langacker:1995eqb,ALEPH:2005ab}, depends on the Weinberg angle or the weak mixing angle and can enable independent determinations of the neutral-current couplings of these fermions. Due to the quark confinement it is relatively difficult to determine the electroweak interactions of quarks. Heavy quarks (c or b) can be determined by tagging the decays of corresponding hadrons containing them. However, light quarks require other flavor separation method. The difficulty in describing the weak interactions of quarks lies in the description of the quark fragmentation process. Thanks to the asymptotic freedom of QCD, the fragmentation process can be studied in the factorization theorem framework  \cite{Collins:1989gx} in the parton model. Factorization theorem tells that measurable quantities, e.g. cross section, can be separated by the calculable hard parts from the non-perturbative soft parts. If only the fragmentation process is taken into consideration, the non-perturbative soft parts are usually factorized as fragmentation functions. Fragmentation functions (FFs) are most important physical quantities in describing hadron productions in high energy reactions. They quantify the hadronization process of quarks and/or gluon in high energy reactions where hadrons are produced. Quantities therefore can be expressed in terms of FFs in the annihilation and other fragmentation processes.

In this paper, we consider the vector meson production in the inclusive electron positron annihilation process and calculate the forward-backward asymmetry in the hadronic level, i.e. the asymmetry in the angular distribution of the produced vector meson. The calculations are carried out in the parton model at leading order twist-4 by applying the collinear expansion formalism~\cite{Ellis:1982wd,Ellis:1982cd,Qiu:1990xxa,Qiu:1990xy}. Collinear expansion is a powerful tool to calculate higher twist effects by taking into account multiple gluon exchange contributions. On the one hand, gauge links will be obtained automatically which make the calculation explicitly gauge invariant. On the other hand, collinear expansion gives a very simple factorization form which consists of calculable hard parts and FFs. This will greatly simplify the systematic calculation of higher twist contributions. After obtaining the differential cross section, we introduce the definition of the forward-backward asymmetry for the production hadron.  We finally present these asymmetry results in terms of FFs. 

The rest of the paper is organized as follows. In Sec.~\ref{sec:definition}, we first introduce the general definition of the forward-backward asymmetry of the muon pair in the electron positron annihilation process and show some conventions used in this paper. 
In Sec.~\ref{sec:formalism}, we present the formalism of vector meson production annihilation process where the differential cross section is given in terms of structure functions. The study in the parton model formulism is given in \ref{sec:partonmodel}, where we present a detailed calculation of how to obtain the hadronic tensor and the cross section at leading order twist-4.
In Sec.~\ref{sec:results}, we present the results for the structure functions and forward-backward asymmetries in terms of the gauge invariant FFs.
A brief summary is given in Sec.~\ref{sec:summary}.

\section{Introduction to the forward-backward asymmetry }\label{sec:definition}

A simple exercise for the fermion pair production in the electron positron annihilation process is to calculate the muon pair production process $e^+(l')+e^-(l)\to \mu^+(k')+\mu^-(k)$. It gives fruitful information about the annihilation reactions. By considering the EW theory, the differential cross section of the this process can be written as
\begin{align}
  \frac{d\sigma}{d\cos\theta}=\frac{ \pi \alpha_{em}^2}{2Q^2}&\bigg\{\chi \left[c_1^ec_1^\mu \left(1+\cos^2\theta \right)+2c_3^ec_3^\mu \cos\theta \right]\nonumber\\
  +& \chi_{int}\left[c_V^ec_V^\mu \left( 1+\cos^2\theta \right)+2c_A^ec_A^\mu \cos\theta \right] \nonumber\\
  +& e_q^2\left(1+\cos^2\theta \right)\bigg\},  \label{f:cross}
\end{align}
where $\theta$ is the scattering angle in the lepton center-of-mass frame or the gauge boson rest frame, $\alpha_{em} = e^2/4\pi$ is the fine structure constant and $Q^2 = q^2=(l+l')^2$.
\begin{align}
   &\chi= \frac{Q^4}{\left[(Q^2-M_Z^2)^2 + \Gamma_Z^2 M_Z^2 \right] \sin^4 2\theta_W}, \\
   &\chi_{int} = -\frac{2e_q Q^2 (Q^2-M_Z^2)}{\left[(Q^2-M_Z^2)^2 + \Gamma_Z^2 M_Z^2 \right] \sin^2 2\theta_W},
\end{align}
where $M_Z$ and  $\Gamma_Z$ are respectively the mass and decay width  of $Z$-boson, $\theta_W$ is the Weinberg angle or the weak mixing angle. $c_1^e = (c_V^e)^2 + (c_A^e)^2$ and $c_3^e = 2 c_V^e c_A^e$,
$c_V^e$ and $c_A^e$ are defined in the weak interaction current
$J_\mu (x)=\bar \psi(x)\Gamma_\mu\psi(x)$ where $\Gamma_\mu= \gamma_\mu (c_V^e - c_A^e \gamma^5)$.
Similar notations are also used for muon and quarks where we use a superscript $\mu$ and $q$ to replace $e$. 

The  forward-backward asymmetry in the angular distribution of positively and negatively charged muons is defined as
\begin{align}
A_{FB} = \frac{\int_{0}^{1} d \sigma_\theta d \cos\theta -\int_{-1}^{0} d \sigma_\theta d \cos\theta  }{\int_{-1}^{1}   d \sigma_\theta d \cos\theta}, \label{f:fbdef}
\end{align}
where $d\sigma_\theta= d \sigma / d\cos\theta$ given in Eq. (\ref{f:cross}). Using the definition in Eq. (\ref{f:fbdef}) and the differential cross section in Eq. (\ref{f:cross}), we have
\begin{align}
A^\mu_{FB} = \frac{3( \chi c_3^e c_3^\mu+\chi_{int} c_A^e c_A^\mu ) }{4(e_q^2+\chi c_1^e c_1^\mu + \chi_{int} c_V^e c_V^\mu  )}. \label{f:fbmnu}
\end{align}
At the low-energy limit ($Q^2<<M_Z^2$), this asymmetry is given approximately by
\begin{align}
 A^\mu_{FB} =- \frac{3G_F Q^2 c_A^e c_A^\mu }{4\sqrt{2}\pi \alpha_{em}},\label{f:fbmu}
\end{align}
where $G_F$ is the Fermi constant. Similar results can also be obtained for quarks as long as we replace the corresponding couplings for muon by that for quarks. For example, as the low-energy limit, we have 
\begin{align}
 A^q_{FB} = \frac{3G_F Q^2 c_A^e c_A^q }{4\sqrt{2}e_q\pi \alpha_{em}}, \label{f:fbq}
\end{align}
where $e_q$ is the electric charge of the quark with flavor $q$.  From Eqs. (\ref{f:fbmnu})-(\ref{f:fbq}), we can see that forward-backward asymmetries depend on weak couplings for certain fermions, they would give independent determinations of these couplings. 
In the following context, we extend the results to the hadron production process.  
We note that the definition of the forward-backward asymmetry in Eq. (\ref{f:fbdef}) will be slightly modified for calculating that for hadrons. It will be shown in Sec. \ref{sec:results}.

\section{The general form of the cross section in terms of structure functions}\label{sec:formalism}

\begin{figure}
  \includegraphics[width=6.0cm]{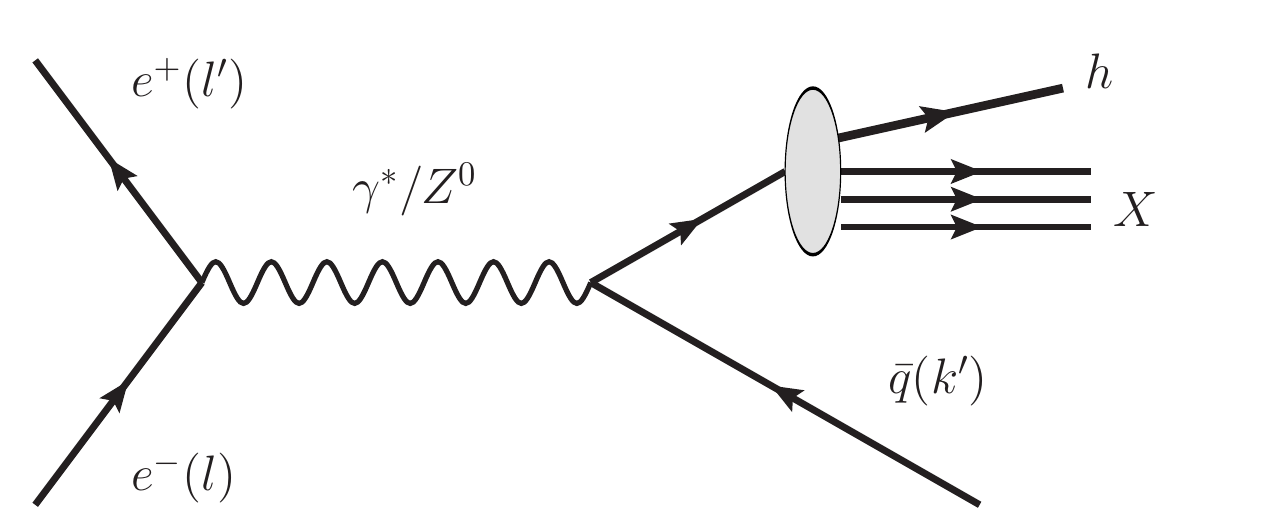}\\
  \caption{Illustrating diagram for the vector boson production inclusive electron positron annihilation process.}\label{anniee}
\end{figure}

At first sight, the vector boson production inclusive electron positron annihilation process can not be calculated because of the lack of perturbative description of the fragmentation process. Therefore, we consider the general decomposition of the hadronic tensor and give the general form of the cross section in terms of structure functions. 
We show this hadron production annihilation process in Fig.~\ref{anniee}. To be explicit, we write down the differential cross section as
\begin{align}
\frac{d\sigma}{dzdy} = \frac{\pi z\alpha_{em}^2}{Q^4}\sum_r \eta_r L_r^{\mu\nu}(l,l') W_{r,\mu\nu}(q,p,S). 
\end{align}
Here we use the notations as illustrated in Fig.~\ref{anniee}. The standard variables  $z=2p\cdot q/Q^2$ and $y=p\cdot l'/p\cdot q$. The subscript $r$ denotes $\gamma\gamma$, $\gamma Z$ and $ZZ$ corresponding respectively to the electromagnetic, interference and weak contributions to this process. 
The summation of $r$ denotes the summation of these cross sections,  i.e. 
\begin{align}
  \frac{d\sigma}{dzdy}=\frac{d\sigma^{\gamma\gamma}}{dzdy}+\frac{d\sigma^{\gamma Z}}{dzdy}+\frac{d\sigma^{ZZ}}{dzdy}. \label{f:crosssum}
\end{align}
Correspondingly, $\eta_{\gamma\gamma} = e_q^2$, $\eta_{\gamma Z} = \chi_{int}$ and $\eta_{ZZ} = \chi$. 

The leptonic tensors for the electromagnetic, interference and weak contributions are respectively given by
\begin{align}
&L^{\gamma\gamma}_{\mu\nu}(l,l') = l_{\mu} l'_{\nu}+l_{\nu}l'_{\mu}-g_{\mu\nu}l\cdot l', \\
&L^{\gamma Z}_{\mu\nu}(l,l') = c_V^e(l_{\mu} l'_{\nu}+l_{\nu}l'_{\mu}-g_{\mu\nu}l\cdot l')+ ic_A^e\varepsilon_{\mu\nu ll'},\\
  &L^{ZZ}_{\mu\nu}(l,l') = c_1^e(l_{\mu} l'_{\nu}+l_{\nu}l'_{\mu}-g_{\mu\nu}l\cdot l')+ ic_3^e\varepsilon_{\mu\nu ll'},
\label{f:leptonictensor}
\end{align}
where $\varepsilon_{\mu\nu AB}\equiv \varepsilon_{\mu\nu\alpha\beta} A^\alpha B^\beta$.
The hadronic tensors are respectively given by
\begin{align}
W^{\gamma\gamma}_{\mu\nu} &(q,p,S) = \frac{1}{2\pi}  \sum_X (2\pi)^4 \delta^4 (q-p- p_X) \nonumber\\
& \times \langle 0| J^{\gamma\gamma}_\nu (0) |p,S;X\rangle \langle p,S;X |J^{\gamma\gamma}_\mu (0)|0\rangle, \\
W^{\gamma Z}_{\mu\nu} &(q,p,S) = \frac{1}{2\pi}  \sum_X (2\pi)^4 \delta^4 (q-p- p_X) \nonumber\\
& \times \langle 0| J^{ZZ}_\nu (0) |p,S;X\rangle \langle p,S;X |J^{\gamma\gamma}_\mu (0)|0\rangle, \\
W^{ZZ}_{\mu\nu} &(q,p,S) = \frac{1}{2\pi}  \sum_X (2\pi)^4 \delta^4 (q-p- p_X) \nonumber\\
& \times \langle 0| J^{ZZ}_\nu (0) |p,S;X\rangle \langle p,S;X |J^{ZZ}_\mu (0)|0\rangle,  
\end{align}
where $S$ denotes the polarization of the hadron and $J^{ZZ}_\mu (x)$ is the quark weak current.

To deal with the hadronic tensor, it is convenient to construct it with known quantities, e.g. momenta.  First of all we show the general decomposition of the hadronic tensor by dividing it into a symmetric and an anti-symmetric part, $W_{\mu\nu} =W^S_{\mu\nu} + iW^A_{\mu\nu}$, where we have omitted the subscript $r=\gamma \gamma, \gamma Z, ZZ$ for simplicity. 
Each of them is given by a linear combination of a set of basic Lorentz tensors (BLTs), i.e.,
\begin{align}
   W^{S\mu\nu} &=\sum_{\sigma,j} W_{\sigma j}^S h_{\sigma j}^{S\mu\nu} + \sum_{\sigma, j} \tilde W_{\sigma j}^S \tilde h_{\sigma j}^{S\mu\nu},\label{f:Wsmunu}\\
   W^{A\mu\nu} &=\sum_{\sigma,j} W_{\sigma j}^A h_{\sigma j}^{A\mu\nu} + \sum_{\sigma, j} \tilde W_{\sigma j}^A \tilde h_{\sigma j}^{A\mu\nu},\label{f:Wamunu}
\end{align}
where $h^{\mu\nu}$ and $\tilde h^{\mu\nu}$ represent the space reflection even and space reflection odd BLTs, respectively. The subscript $\sigma$ specifies the polarization. In the general decomposition of the hadronic tensor, we require  that hadron tensors corresponding to the electromagnetic, interference and weak contributions have the same form. Therefore, they can by summed together.

For inclusive reactions, the unpolarized or the spin-independent BLTs can only be constructed by momentum vectors, $q, p$.
There are in total 3 unpolarized BLTs given by
\begin{align}
   h^{S\mu\nu}_{Ui}&=\Big\{g^{\mu\nu}-\frac{q^\mu q^\nu}{q^2}, ~p_q^\mu  p_q^\nu \Big\},\label{f:hsU} \\
   \tilde h^{A\mu\nu}_{U}&=\varepsilon ^{\mu\nu qp}.\label{f:thaU}
\end{align}
The subscript $U$ denotes the unpolarized part.
Here $p_q^\mu\equiv p^\mu- q^\mu(p\cdot q)/q^2$ which satisfies $p_q\cdot q=0$. This notation ensures that the hadron tensor satisfies the current conservation.

The vector polarization dependent BLTs are given by
\begin{align}
  h^{S\mu\nu}_{Vi} &=\varepsilon^{\{\mu qpS}p_q^{\nu\}}, \label{f:hsV}\\
 \tilde h^{S\mu\nu}_{Vi} &=\Big\{(q\cdot S)h^{S\mu\nu}_{Ui}, ~ S_q^{\{\mu}p_q^{\nu\}} \Big\}, \label{f:thsV}\\
  h^{A\mu\nu}_{Vi} &=\Big\{(q\cdot S)\tilde h^{A\mu\nu}_{U}, ~ \varepsilon^{[\mu qpS}p_q^{\nu]} \Big\},\label{f:haV}\\
  \tilde h^{A\mu\nu}_{Vi} &=S_q^{[\mu}p_q^{\nu]} ,\label{f:thaV}
\end{align}
where $A^{\{\mu}B^{\nu\}} \equiv A^\mu B^\nu +A^\nu B^\mu$  and $A^{[\mu}B^{\nu]} \equiv A^\mu B^\nu -A^\nu B^\mu$.
There are 7 such vector polarized BLTs in total.

The tensor polarized part is composed of $S_{LL}$-, $S_{LT}$- and $S_{TT}$-dependent parts. More discussions of polarizations of spin one particles can be found in ref. \cite{Bacchetta:2000jk}. For simplicity, we do not show them in this paper.  The tensor polarized part can be taken as a product of the unpolarized BLTs and polarization dependent Lorentz scalar(s) or pseudo-scalar(s). 
They are given by
\begin{align}
  h_{LLi}^{S\mu\nu}&=S_{LL} h^{S\mu\nu}_{Ui}, \label{f:hsLL} \\
  \tilde h_{LL}^{A\mu\nu}&=S_{LL} \tilde h^{A\mu\nu}_{U}, \label{f:thsLL} \\
  h^{S\mu\nu}_{LT} &=S_{LT}^{\{\mu}p_q^{\nu\}}, \label{f:hsLT}\\
 \tilde h^{S\mu\nu}_{LT} &=\varepsilon^{\{\mu qp S_{LT}}p_q^{\nu\}}, \label{f:thsLT}\\
  h^{A\mu\nu}_{LT} &=S_{LT}^{[\mu}p_q^{\nu]},\label{f:haLT}\\
  \tilde h^{A\mu\nu}_{LT} &=\varepsilon^{[\mu qp S_{LT}}p_q^{\nu]} ,\label{f:thaLT} \\
  h^{S\mu\nu}_{TT} &=S_{TT}^{\mu\nu}, \label{f:hsTT}\\
  \tilde h^{S\mu\nu}_{TT} &=\varepsilon^{\{\mu \alpha qp}S_{TT}^{\nu\}\alpha}. \label{f:thsTT}
\end{align}

Substituting Eqs. (\ref{f:hsU})-(\ref{f:thsTT}) into Eqs. (\ref{f:Wsmunu})-(\ref{f:Wamunu}) and contracting with the leptonic tensor yield the differential cross section. The forward-backward asymmetry is defined in the lepton center-of-mass frame.  To be consistent with the definition, we show this cross section in the same frame  in which
\begin{align}
&p=(E_p,0,0,p_z), \\
&l=Q(1,\sin\theta,0,\cos\theta)/2, \\
&S=(\lambda_h\frac{p_z}{M},|S_T|\cos\varphi_S, |S_T|\sin\varphi_S,\lambda_h\frac{E_P}{M}), \\
&S_{LT}=(0,|S_{LT}|\cos\varphi_{LT}, |S_{LT}|\sin\varphi_{LT},0), \\
&S_{TT}^{x\mu}= (0,|S_{TT}|\cos2\varphi_{TT}, |S_{TT}|\sin2\varphi_{TT},0).
\end{align}
After making Lorentz contraction with the leptonic tensor, we obtain the general form for the cross section,
\begin{align}
  &\frac{d\sigma}{dzdy} =\frac{\pi z\alpha_{em}^2}{Q^2}\sum_r \eta_r \Bigl[
  \mathcal{F}_U +S_{LL}\mathcal{F}_{LL}+|S_T|(\mathcal{F}_T + \tilde{\mathcal{F}}_T)\nonumber\\
 &+\lambda_h \tilde{\mathcal{F}}_L +|S_{LT}|(\mathcal{F}_{LT}+ \tilde{\mathcal{F}}_{LT})+|S_{TT}|(\mathcal{F}_{TT}+ \tilde{\mathcal{F}}_{TT})
  \Bigr], \label{f:csinclusive}
\end{align}
where we use $\mathcal{F}$ and $\tilde{\mathcal{F}}$ to denote the parity conserved and parity violated parts, respectively. These explicit expressions are given by
\begin{align}
&{\cal F}_{U} =2A(y) F_{U1}+  D^2(y) F_{U2} + B(y) F_{U3},\label{f:FUin} \\
&\tilde{\cal F}_{L} = 2A(y) \tilde F_{L1}+ D^2(y) \tilde F_{L2} + B(y) \tilde F_{L3}, \label{f:tFLin} \\
&{\cal F}_{LL}= 2A(y) F_{LL 1}+ D^2(y) F_{LL2} + B(y) F_{LL3}, \label{f:FLLin} \\
&{\cal F}_{T}= \sin\varphi_S \left[ D(y) F_{T1}^{\sin\varphi_S} + 2C(y) F_{T2}^{\sin\varphi_S} \right], \label{f:FTin}\\
&\tilde{\cal F}_{T}= \cos\varphi_S \left[ D(y) \tilde F_{T1}^{\cos\varphi_S} + 2C(y) \tilde F_{T2}^{\cos\varphi_S} \right], \label{f:tFTin} \\
&{\cal F}_{LT}= \cos\varphi_{LT} \left[ D(y) F_{LT1}^{\cos\varphi_{LT}} + 2C(y) F_{LT2}^{\cos\varphi_{LT}} \right], \label{f:FLTin}\\
&\tilde{\cal F}_{LT} = \sin\varphi_{LT} \left[ D(y) \tilde F_{LT1}^{\sin\varphi_{LT}} + 2C(y) \tilde F_{LT2}^{\sin\varphi_{LT}} \right], \label{f:tFLTin} \\
&{\cal F}_{TT}= \cos2\varphi_{TT} D^2(y) F_{TT}^{\cos2\varphi_{TT}}, \label{f:FTTin} \\
&\tilde{\cal F}_{TT}= \sin2\varphi_{TT} D^2(y) \tilde F_{TT}^{\sin2\varphi_{TT}}, \label{f:tFTTin}
\end{align}
where $F$ and $\tilde F$ with subscripts $U, L, LL,T, LT$ and $TT$ are known as structure functions. We have in total 19 inclusive structure functions.
Here we have defined 
\begin{align}
 & A(y)=(1-y)^2+y^2=\frac{1}{2}(1+\cos^2\theta), \\
 & B(y)=2y-1=\cos\theta, \\
 & C(y)=2(2y-1)\sqrt{y(1-y)}=\frac{1}{2}\sin2\theta, \\
 & D(y)=2\sqrt{y(1-y)}=\sin\theta, 
\end{align}
with $y=(1+\cos\theta)/2$.
We can see that all the $\theta$ dependent terms are given explicitly. 
From Eq. (\ref{f:cross}) we see that the differential cross section at the quark level depends on $A(y)$ and $B(y)$ only. This implies that $F_{U1,3}, \tilde F_{L1,3}$ and $F_{LL1,3}$ are leading twist structure functions (may have higher twist corrections) while the other terms are higher twist ones.

\section{The cross section in the parton model} \label{sec:partonmodel}

As mentioned in the $introduction$, the difficulty in describing the weak interactions of quarks lies in the description of the fragmentation process. The parton model which is applicable to any hadronic cross section involving a large momentum transfer can be used to describe this. Measurable quantity is then factorized as a convolution of the hard part and the non-perturbative soft part.  If only the fragmentation process is taken into consideration, the non-perturbative soft parts are usually factorized as fragmentation functions. This is the case in this paper. 
In this section, in the parton model framework,  we present a detailed calculation of how to obtain the hadronic tensor and the cross section at leading order twist-4.

\subsection{The general forms of hadronic tensors in the parton model}

\begin{figure}
 \includegraphics[width=7cm]{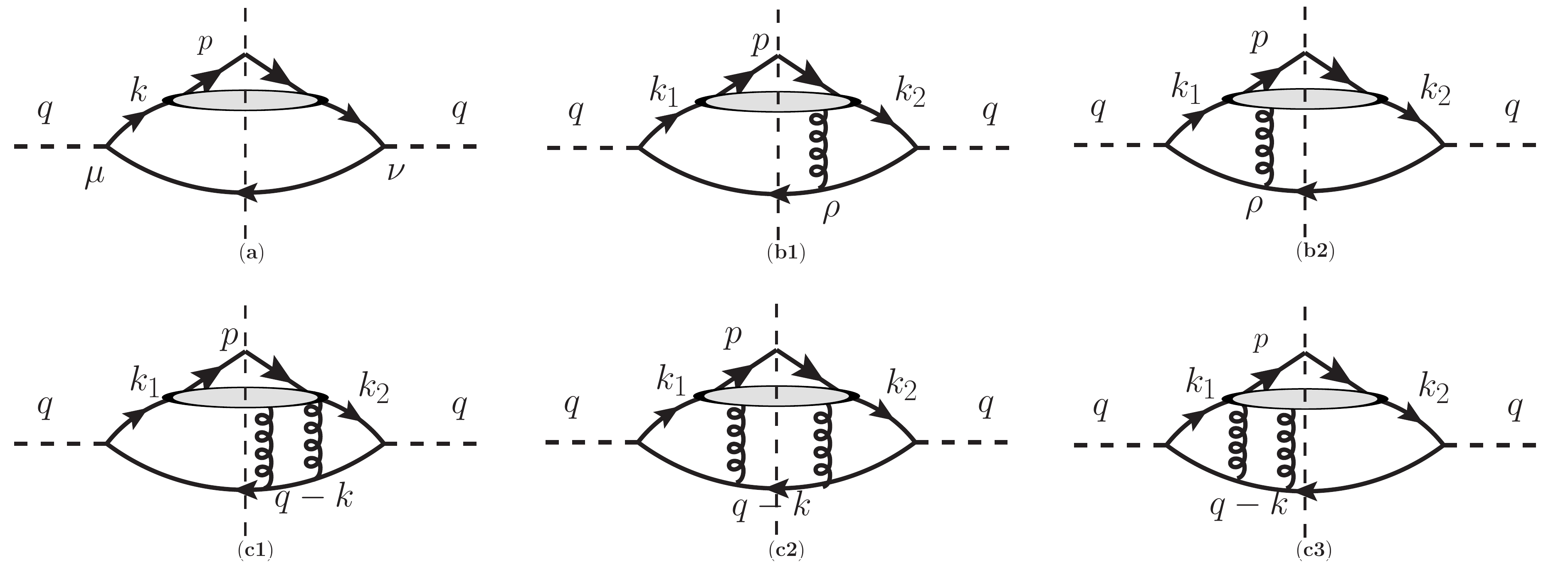}
  \caption{The first few diagrams as examples of the considered diagram series with exchange of $j$-gluon(s) and different cuts.
 We see (a) $j=0$, (b1) $j=1$ and left cut, (b2) $j=1$ and right cut,
  (c1)  $j=2$ and left cut, (c2) $j=2$ and middle cut,  and (c3) $j=2$ and right cut, respectively.} \label{fig:Feyndiagram}
\end{figure}

The parton model gives a reliable approximate description of the hadronic interactions and an opportunity to calculate the forward-backward asymmetry at hadronic level. By applying this model, we limit our considerations at the tree level or leading order of the QCD and consider the series of diagrams illustrated in Fig. \ref{fig:Feyndiagram}, where diagrams with exchange of $j$ gluon(s) ($j=0,1,2, \cdots$) are included.

To obtain the explicit expression of the hadronic tensor, we use the collinear expansion formalism. It provides not only the correct formalism where the differential cross section or the hadronic tensor is given in terms of gauge invariant FFs, but also very simplified expressions so that even twist-4 contributions can be calculated. 
After the collinear expansion, the hadronic tensor is obtained as \cite{Wei:2013csa,Wei:2014pma,Yang:2017sxz,Yang:2020ezt}
\begin{align}
&W_{\mu\nu}(q,p,S)=\sum_{j,c}\tilde W^{(j,c)}_{\mu\nu}(q,p,S),\label{f:Wsum}
\end{align}
where $c=L, R, M$ denotes different cuts for left, right and middle, respectively.
The  $\tilde W^{(j,c)}_{\mu\nu}$ is a trace of the collinear-expanded hard part and gauge invariant quark-$j$-gluon-quark correlator. In other words, the hadronic tensor is written as 
an explicit factorized form. To be explicit, we have
\begin{align}
   & \tilde W^{(0)}_{\mu\nu}
   = \frac{1}{2} \mathrm{Tr}\big[\hat h^{(0)}_{\mu\nu} \hat\Xi^{(0)}\big],\label{f:tW0}\\
   & \tilde W^{(1,L)}_{\mu\nu}
   =-\frac{1}{4(p\cdot q)} \mathrm{Tr}\big[\hat h^{(1)\rho}_{\mu\nu}\hat\Xi^{(1)}_{\rho} \big],\label{f:tW1L}\\
   & \tilde W^{(2,M)}_{\mu\nu}
   =\frac{1}{4(p\cdot q)^2} \mathrm{Tr}\big[\hat h^{(2)\rho\sigma}_{\mu\nu} \hat\Xi^{(2,M)}_{\rho\sigma} \big],\label{f:tW2M}\\
   & \tilde W^{(2,L)}_{\mu\nu}
   =\frac{1}{4(p\cdot q)^2} \mathrm{Tr}\big[ \hat N_{\mu\nu}^{(2)\rho\sigma} \hat\Xi^{(2)}_{\rho\sigma} +\hat h^{(1)\rho}_{\mu\nu}\hat\Xi^{(2\prime)}_{\rho} \big],\label{f:tW2L}
\end{align}
where we have omitted the arguments for simplicity. The hard parts or simplified scattering amplitudes are given by
\begin{align}
&\hat h^{(0)}_{\mu\nu} = \Gamma_\mu^q \slashed n \Gamma_\nu^q/p^+ , \label{f:h0}\\
&\hat h_{\mu\nu}^{(1)\rho} = \Gamma_\mu^q \slashed n \gamma^\rho \slashed{\bar n}\Gamma_\nu^q,  \label{f:h1}\\
&\hat N^{(2)\rho\sigma}_{\mu\nu} = q^- \Gamma_\mu \gamma^\rho \slashed n \gamma^\sigma \Gamma_\nu, \label{f:N2}\\
&\hat h_{\mu\nu}^{(2)\rho\sigma}=p^+ \Gamma_\mu \slashed{\bar n} \gamma^\rho \slashed n \gamma^\sigma \slashed{\bar n} \Gamma_\nu/2.  \label{f:h2}
\end{align}
The corresponding quark-$j$-gluon-quark correlators are given by
\begin{align}
\hat \Xi^{(0)} =& \sum_X\int \frac{ p^+d \xi^-}{2\pi} e^{-ik^+\xi^-}
  \langle 0 | \mathcal{L}^\dagger(0,\infty)\psi(0) |hX\rangle\nonumber\\
&\times \langle hX| \bar\psi(\xi^-) \mathcal{L}(\xi^-,\infty) |0\rangle,\label{f:Xi0}\\
  \hat \Xi^{(1)}_{\rho}  = & \sum_X \int\frac{p^+ d\xi^-}{2\pi} e^{-ip^+\xi^-/z}\langle 0 | \mathcal{L}^\dagger (0,\infty) D_{\rho}(0) \psi(0) |hX\rangle \nonumber\\
&\times\langle hX| \bar\psi(\xi^-) \mathcal{L}(\xi^-,\infty) |0\rangle, \label{f:t3Xi1} \\
\hat \Xi^{(2M)}_{\rho\sigma} = & \sum_X \int\frac{p^+d\xi^-}{2\pi} e^{-ip^+\xi^-/z} \langle 0 | \mathcal{L}^\dagger (0,\infty) D_{\rho}(0) \psi(0) |hX\rangle
\nonumber\\
& \times\langle hX| \bar\psi(\xi^-) D_\sigma (\xi^-)  \mathcal{L}(\xi^-,\infty) |0\rangle, \\
\hat \Xi^{(2')}_{\rho} = & \sum_X \int\frac{p^+d\xi^-}{2\pi} e^{-ip^+\xi^-/z}  p^\sigma \langle 0 | \mathcal{L}^\dagger (0,\infty) D_{\rho}(0)  D_\sigma (0)\nonumber\\
&\times \psi(0) |hX\rangle\langle hX|\bar\psi(\xi^-) \mathcal{L}(\xi^-,\infty) |0\rangle, \label{f:Xi2B}\\
 \hat \Xi^{(2)}_{\rho\sigma}= & \sum_X \int\frac{p^+ d\xi^- }{2\pi} ip^+d\eta^- e^{-ip^+\xi^-/z}e^{-ip^+\eta^-/z} \langle 0 |
 \nonumber \\
 &\times \mathcal{L}^\dagger (\eta^-,\infty) D_{\rho}(\eta^-) D_{\sigma} (\eta^-)\mathcal{L}^\dagger (0,\eta^-)  \psi(0) |hX\rangle \nonumber\\
 &\times \langle hX| \bar\psi(\xi^-) \mathcal{L}(\xi^-,\infty) |0\rangle. \label{f:Xi2M}
\end{align}
where $D_\rho=-i\partial_\rho+gA_\rho$ are the transverse covariant derivative, $\mathcal{L}(0,y)$ is the gauge link.
The argument $\xi$ in the quark filed operator $\psi$ and gauge link represents $(0,\xi^-)$.
We note that the leading power contribution of $\tilde W^{(j)}_{\mu\nu}$ is twist-$(j+2)$.
Therefore the second term in Eq.~(\ref{f:tW2L}) has no contribution up to twist-4 because of of the factor $p^\sigma$ in the definition of $\hat\Xi^{(2')}_\rho$ given by Eq.~(\ref{f:Xi2B}).
The leading power contribution of this term is twist-5.

\subsection{Decompositions of correlators}\label{sec:decomp}

In the previous part, hadronic tensors are given in the explicit factorization forms where hard parts and non-perturbative soft parts are naturally separated. These soft parts are correlators shown in Eqs. (\ref{f:Xi0})-(\ref{f:Xi2M}). 
Correlators can not be calculated with perturbative theory because they contain the hadronization information. However,  they are $4\times 4$ matrices in Dirac space and can be decomposed in terms of $\Gamma$ matrices, i.e., $\Gamma = \{I, i\gamma^5, \gamma^\alpha, \gamma^5\gamma^\alpha, i\sigma^{\alpha\beta}\gamma^5 \}$. The decomposition can be written explicitly as
\begin{align}
  \hat\Xi=I\Xi+i\gamma^5\tilde\Xi +\gamma^\alpha \Xi_\alpha + \gamma^5 \gamma^\alpha \tilde\Xi_\alpha
  + i\sigma^{\alpha\beta}\gamma^5\Xi_{\alpha\beta}. \label{f:XiD}
\end{align}
In the inclusive electron positron annihilation process, only the chiral even quantities are involved due to no spin flip. Thus we only need to consider the $\gamma^\alpha$- and the $\gamma^5\gamma^\alpha$-term in the decomposition
of the correlators in terms of the $\Gamma$-matrices and corresponding coefficient functions, such as $\hat\Xi^{(0)} =\Xi_\alpha^{(0)}\gamma^\alpha+\tilde\Xi_\alpha^{(0)} \gamma^5\gamma^\alpha+\cdots$.

We first consider the quark-quark correlator $\hat\Xi^{(0)}$ . At twist-4,  the coefficient functions are given by
\begin{align}
  z\Xi^{(0)}_\alpha =&p^+\bar{n}_\alpha\big(D_1+S_{LL}D_{1LL}  \big) - M\tilde S_{T\alpha}D_T \nonumber\\
  & -M S_{LT\alpha}D_{LT} +\frac{M^2}{p^+}n_\alpha \big(D_3 + S_{LL}D_{3LL}\big),\label{f:xi0even}\\
  z\tilde\Xi^{(0)}_\alpha =&-p^+\bar{n}_\alpha\lambda_hG_{1L}- M S_{T\alpha}G_T \nonumber\\
  & -M \tilde S_{LT\alpha}G_{LT}-\frac{M^2}{p^+}n_\alpha \lambda_hG_{3L} .\label{f:xi0odd}
\end{align}
Here $\tilde S_{T\alpha}=\varepsilon_{\perp \beta\alpha}S_T^\beta$.  $D$'s and $G$'s represent the $\gamma^\alpha$- and $\gamma^5\gamma^\alpha$-type FFs, respectively.
The digit $j$ in the subscript denotes twist-$(j+1)$;
the capital letter such as $L$, $T$, $LL$ and $LT$ denote hadron polarizations.

For the quark-gluon-quark correlator $\hat\Xi^{(1)}_\rho $, the chiral even parts are
\begin{align}
  z\Xi^{(1)}_{\rho\alpha}=&p^+\bar n_\alpha M\big(\tilde S_{T\rho}D_{dT}+S_{LT\rho} D_{dLT}\big)\nonumber\\
  &+ M^2 g_{\perp\rho\alpha} \big(D_{3d}+S_{LL}D_{3dLL}\big)\nonumber\\
  &+ i M^2\varepsilon_{\perp\rho\alpha}\lambda_h D_{3dL} , \label{f:xi1even3} \\
  z\tilde\Xi^{(1)}_{\rho\alpha}=&ip^+\bar n_\alpha M\big( S_{T\rho}G_{dT} +\tilde S_{LT\rho} G_{dLT}\big)\nonumber\\
  &+iM^2\varepsilon_{\perp\rho\alpha} \big(G_{3d}+S_{LL}G_{3dLL}\big) \nonumber\\
  &+ M^2 g_{\perp\rho\alpha}\lambda_h G_{3dL}, \label{f:xi1odd3}
\end{align}
where we use subscript $d$ to denote FFs which are defined through $\hat \Xi^{(1)}_\rho$.

For the quark-gluon-gluon-quark correlators $\hat\Xi^{(2)}_{\rho\sigma}$ and $\hat\Xi^{(2,M)}_{\rho\sigma}$,  we require that the decomposition of $\hat\Xi^{(2,M)}_{\rho\sigma} $ takes exactly the same form as that of $\hat\Xi^{(2)}_{\rho\sigma}$. We just add an additional superscript $M$ to distinguish them in the following context from each other. For  the chiral even part, the corresponding coefficient function are given by
\begin{align}
 z\Xi^{(2)}_{\rho\sigma\alpha} = & p^+\bar n_\alpha M^2\Big[
   g_{\perp\rho\sigma} D_{3dd}+ i \varepsilon_{\perp\rho\sigma} \lambda_h D_{3dd L} 
   \Big], \label{f:Xi2decomp} \\
 z\tilde\Xi^{(2)}_{\rho\sigma\alpha} =& p^+\bar n_\alpha M^2\big[
    i \varepsilon_{\perp\rho\sigma} G_{3dd}+ M^2 g_{\perp\rho\sigma} \lambda_h G_{3dd L}\big], \label{f:tXi2decomp}
\end{align}
where  we use $dd$ in the subscript to denote FFs defined via $\hat\Xi^{(2)}_{\rho\sigma}$.
From Eqs.~(\ref{f:xi0even})-(\ref{f:tXi2decomp}), we see that for the twist-4 parts, the decomposition of $\Xi$ and that of $\tilde \Xi$ have exact one to one correspondence. For each $D_3$, there is a $G_3$ corresponding to it.
They always appear in pairs. Because of the Hermiticity of $\hat\Xi^{(0)}$ and $\hat\Xi^{(2,M)}_{\rho\sigma}$, FFs defined via them are real. For those defined via $\hat\Xi^{(1)}_{\rho}$ and $\hat\Xi^{(2)}_{\rho\sigma}$, there is no such constraint so that they can be complex.

Not all the FFs defined in Eqs.~(\ref{f:xi0even})-(\ref{f:tXi2decomp}) are actually independent. We can eliminate the correlated terms by using the QCD equation of motion $\gamma\cdot D\psi=0$. With this equation, the quark-$j$-gluon-quark correlators is related to the quark-quark correlator. For the two transverse components $\Xi_{\perp}^{(0)\rho}$ and $\tilde\Xi_{\perp}^{(0)\rho}$, we have
\begin{align}
  k^+\Xi_{\perp}^{(0)\rho}&=-g_\perp^{\rho\sigma} \mathrm{Re}\Xi^{(1)}_{\sigma +}-\varepsilon^{\rho\sigma}_{\perp }\mathrm{ Im} \tilde \Xi^{(1)}_{\sigma +},\label{f:perpeom}\\
  k^+\tilde \Xi_{\perp}^{(0)\rho}&=-g_\perp^{\rho\sigma} \mathrm{Re}\tilde \Xi^{(1)}_{\sigma +}-\varepsilon^{\rho\sigma}_{\perp }\mathrm{ Im} \Xi^{(1)}_{\sigma +}. \label{f:tperpeom}
\end{align}
Equations (\ref{f:perpeom}) and (\ref{f:tperpeom}) lead to a set of relationships between twist-3 FFs which can be given in the unified form
\begin{align}
  D_{S} - iG_{S} =-z(D_{dS}-G_{dS}), \label{f:t3eom}
\end{align}
where $S=$ $T$ and $LT$ whenever applicable.
Similarly, for the minus components of $\Xi^{(0)}_\alpha$ and $\tilde\Xi^{(0)}_\alpha$, we have
\begin{align}
  2k^{+2}\Xi^{(0)}_- &=k^+\Big(g_\perp^{\rho\sigma} \Xi^{(1)}_{\rho\sigma } + i\varepsilon_\perp^{\rho\sigma}\tilde \Xi^{(1)}_{\rho\sigma }\Big) \nonumber\\
  & =- g_\perp^{\rho\sigma} \Xi^{(2,M)}_{\rho\sigma +}+i\varepsilon_\perp^{\rho\sigma}\tilde \Xi^{(2,M)}_{\rho\sigma +},\label{eq:minuseom}\\
  2k^{+2}\tilde\Xi^{(0)}_- &=k^+\Big(g_\perp^{\rho\sigma}\tilde \Xi^{(1)}_{\rho\sigma } + i\varepsilon_\perp^{\rho\sigma} \Xi^{(1)}_{\rho\sigma }\Big) \nonumber\\
  & =- g_\perp^{\rho\sigma}\tilde \Xi^{(2,M)}_{\rho\sigma +}+i\varepsilon_\perp^{\rho\sigma}\Xi^{(2,M)}_{\rho\sigma +}. \label{eq:tminuseom}
\end{align}
From Eqs.~(\ref{eq:minuseom}) and (\ref{eq:tminuseom}), we obtain a set of relationships between twist-4 FFs
defined via $\hat\Xi^{(0)}$, $\hat\Xi^{(1)}$ and $\hat\Xi^{(2,M)}$,
\begin{align}
&D_{3}  = z D_{-3d} = - z^2D_{+3dd}^{M}, \label{f:t4eomD} \\
&D_{3LL}  = z D_{-3dLL} = - z^2D_{+3ddLL}^{M}, \label{f:t4eomDLL} \\
&G_{3L}  = z D_{-3dL} = z^2D_{+3ddL}^{M}, \label{f:t4eomGL}
\end{align}
where $D_{\pm}\equiv D\pm G$ such as $D_{-3d}\equiv D_{3d}-G_{3d}$ and so on. We note here that only the one-dimensional or collinear FFs are shown. In fact the relationships between three-dimensional or transverse momentum dependent FFs can be obtained in the same way \cite{Chen:2016moq}, we do not repeat them in this paper.

\subsection{The hadronic tensor at twist-4}

It is straightforward to calculate the hadronic tensor with FFs and the corresponding hard parts at hand. The important step is calculating these traces.
To be explicit, we calculate the leading twist, twist-3 and twist-4 contributions in turn. 
The leading twist contributions only come from the quark-quark correlator $\Xi^{(0)}$. The corresponding traces are simple and given by
\begin{align}
  & \mathrm{Tr}\big[\hat h^{(0)}_{\mu\nu} \slashed {\bar n}\big]
 =-\frac{4}{p^+}\big( c_1^qg_{\perp\mu\nu}+ic_3^q\varepsilon_{\perp\mu\nu}\big),\label{f:traceh0} \\
  & \mathrm{Tr}\big[\hat h^{(0)}_{\mu\nu} \gamma^5 \slashed {\bar n}\big]
 =\frac{4}{p^+}\big( c_3^qg_{\perp\mu\nu}+ic_1^q\varepsilon_{\perp\mu\nu}\big). \label{f:traceh05}
\end{align}
And the leading twist hadronic tensor is 
\begin{align}
  z\tilde W_{t2\mu\nu} =& -2\Big[c_1^qg_{\perp\mu\nu}+ic_3^q\varepsilon_{\perp\mu\nu}\Big]\big( D_1+S_{LL}D_{1LL}\big) \nonumber\\
  &-2\Big[c_3^qg_{\perp\mu\nu}+ic_1^q\varepsilon_{\perp\mu\nu}\Big]\lambda_h G_{1L}. \label{f:Wt2munu}
\end{align}
We find that $\tilde W_{t2\mu\nu}$ satisfies the current conservation, i.e. $q^\mu \tilde W_{t2\mu\nu}=q^\nu \tilde W_{t2\mu\nu}=0$.

Twist-3 contributions have two origins, one is the quark-quark correlator $\hat \Xi^{(0)}$ and the other is the quark-gluon-quark correlator $\hat \Xi^{(1)}_\rho$. We first calculate these contributions from the quark-quark correlator $\hat \Xi^{(0)}$. Here we use traces 
\begin{align}
  & \mathrm{Tr}\big[\hat h^{(0)}_{\mu\nu} \slashed {k}\big]
 =\frac{4}{p^+}\big( c_1^qk_{\{\mu} n_{\nu\}}+ic_3^q \tilde k_{[\mu} n_{\nu]} \big),\label{f:tracet3h0} \\
  & \mathrm{Tr}\big[\hat h^{(0)}_{\mu\nu} \gamma^5 \slashed {k}\big]
 =-\frac{4}{p^+}\big( c_3^qk_{\{\mu} n_{\nu\}}+ic_1^q \tilde k_{[\mu} n_{\nu]} \big),\label{f:traceht305}
\end{align}
where $k$ denote $S_T (\tilde S_T)$ and $S_{LT} (\tilde S_{LT})$  and obtain
\begin{align}
  z\tilde W^{(0)}_{t3\mu\nu} = -\frac{2M}{p^+}&\Big[\big(c_1^q \tilde S_{T\{\mu}n_{\nu\}}-ic_3^q S_{T[\mu}n_{\nu]}\big)D_T \nonumber\\
  +&\big(c_1^q S_{LT\{\mu}n_{\nu\}}+ic_3^q \tilde S_{LT[\mu}n_{\nu]}\big)D_{LT}\Big] \nonumber\\
  +\frac{2M}{p^+}&\Big[\big(c_3^q S_{T\{\mu}n_{\nu\}}+ic_1^q \tilde S_{T[\mu}n_{\nu]}\big)G_T \nonumber\\
  +&\big(c_3^q \tilde S_{LT\{\mu}n_{\nu\}}-ic_1^q S_{LT[\mu}n_{\nu]}\big)G_{LT}\Big]. \label{f:W0t3}
\end{align}
For the twist-3 contribution from the quark-gluon-quark correlator $\hat \Xi^{(1)}_\rho$, we use
\begin{align}
 & \mathrm{Tr}\big[\hat h^{(1)\rho}_{\mu\nu} \slashed{\bar n}\big]
 = -8\big(c_1^q g_{\perp\mu}^{\rho} \bar n^\nu +ic_3^q \varepsilon_{\perp\mu}^\rho\bar n^\nu \big), \label{f:tracet3h1}\\
 & \mathrm{Tr}\big[\hat h^{(1)\rho}_{\mu\nu}\gamma^5 \slashed{\bar n}\big]
  =+8\big(c_3^q g_{\perp\mu}^{\rho} \bar n^\nu +ic_1^q \varepsilon_{\perp\mu}^\rho\bar n^\nu \big), \label{f:traceht3h51}
\end{align}
and obtain
\begin{align}
  z\tilde W^{(1)L}_{t3\mu\nu} = \frac{2p^+M}{p\cdot q}& \Big[\big(c_1^q \tilde S_{T\{\mu} \bar n_{\nu\}}-ic_3^q S_{T[\mu} \bar n_{\nu ]}\big)D_{dT}\nonumber\\
  +&\big(c_1^q S_{LT\{\mu} \bar n_{\nu\}}+ic_3^q \tilde S_{LT[\mu} \bar n_{\nu ]}\big)D_{dLT} \Big]\nonumber\\
  -\frac{2p^+M}{p\cdot q}& \Big[\big(c_3^q \tilde S_{T\{\mu} \bar n_{\nu\}}+ic_1^q S_{T[\mu} \bar n_{\nu ]}\big)G_{dT}\nonumber\\
  +&\big(c_3^q S_{LT\{\mu} \bar n_{\nu\}}-ic_1^q \tilde S_{LT[\mu} \bar n_{\nu ]}\big)G_{dLT}\Big]. \label{f:W1t3Ld}
\end{align}

The complete twist-3 hadronic tensor is the sum of all the twist-3 contributions, i.e, $\tilde W_{t3\mu\nu}=\tilde W^{(0)}_{t3\mu\nu}+\tilde W^{(1)L}_{t3\mu\nu}+\left(\tilde W^{(1)L}_{t3\nu\mu}\right)^*$.
Using Eqs. (\ref{f:t3eom}), (\ref{f:W0t3}) and (\ref{f:W1t3Ld}), we eliminate the non-independent FFs and obtain the complete hadronic tensor at twist-3.
\begin{align}
  z\tilde W_{t3\mu\nu} = -\frac{2M}{p\cdot q}&\Big[\big(c_1^q \tilde S_{T\{\mu}\bar q_{\nu\}}-ic_3^q S_{T[\mu}\bar q_{\nu]}\big)D_T \nonumber\\
  +&\big(c_1^q S_{LT\{\mu}\bar q_{\nu\}}+ic_3^q \tilde S_{LT[\mu}\bar q_{\nu]}\big)D_{LT}\Big] \nonumber\\
  +\frac{2M}{p\cdot q}&\Big[\big(c_3^q S_{T\{\mu}\bar q_{\nu\}}+ic_1^q \tilde S_{T[\mu}\bar q_{\nu]}\big)G_T \nonumber\\
  +&\big(c_3^q \tilde S_{LT\{\mu}\bar q_{\nu\}}-ic_1^q S_{LT[\mu}\bar q_{\nu]}\big)G_{LT}\Big]. \label{f:Wt3}
\end{align}
where $\bar q =q -2p/z$. It it can be shown that $\tilde W_{t3\mu\nu}$ satisfies the current conservation $q^\mu \tilde W_{t3\mu\nu}=q^\nu \tilde W_{t3\mu\nu}=0$. Here we can see that consideration of the quark(-j)-gluon-quark correlator is also a requirement of the current conservation.

As for the twist-4 contributions, they have three origins which correspond to correlators $\hat \Xi^{(0)}$, $\hat \Xi^{(1)}_\rho$ and $\hat \Xi^{(2)}_{\rho\sigma}$, respectively.  We first calculate contributions from quark-quark correlator $\hat \Xi^{(0)}$.   Using the following two traces, 
\begin{align}
  & \mathrm{Tr}\big[\hat h^{(0)}_{\mu\nu} \slashed n\big]
 =\frac{8}{p^+}c_1^qn_\mu n_\nu,\label{f:traceh0t4} \\
  & \mathrm{Tr}\big[\hat h^{(0)}_{\mu\nu} \gamma^5 \slashed n\big]
 =-\frac{8}{p^+}c_3^qn_\mu n_\nu,\label{f:traceh05t4}
\end{align}
we obtain
\begin{align}
  z\tilde W^{(0)}_{t4\mu\nu} =& -\frac{4M^2}{(p^+)^2}c_1^q n_\mu n_\nu\big( D_3+S_{LL}D_{3LL}\big) \nonumber\\
  &+ \frac{4M^2}{(p^+)^2}c_3^qn_\mu n_\nu\lambda_h G_{3L}. \label{f:W0t4munu}
\end{align}

For the twist-4 contributions from quark-gluon-quark correlator $\hat \Xi^{(1)}_\rho$, we use
\begin{align}
  & \mathrm{Tr}\big[\hat h^{(1)}_{\mu\nu} \gamma^\alpha \big](g_{\perp\rho\alpha} +i\varepsilon_{\perp\rho\alpha} )
 =\frac{16}{p^+}\big(c_1^q  +c_3^q \big)n_\mu \bar n_\nu,\label{f:traceh1t4} \\
  & \mathrm{Tr}\big[\hat h^{(1)}_{\mu\nu} \gamma^5 \gamma^\alpha \big](i\varepsilon_{\perp\rho\alpha}+g_{\perp\rho\alpha})
 =-\frac{16}{p^+}\big(c_1^q  +c_3^q \big) n_\mu \bar n_\nu ,\label{f:traceh15t4}
\end{align}
and obtain
\begin{align}
  z\tilde W^{(1)L}_{t4\mu\nu} = -\frac{4M^2}{p\cdot q} n_\mu n_\nu
  &\Big[c_1^q\big(D_{3d}+S_{LL}D_{3dLL}\big) +c_3^q \lambda_h D_{3dL}\Big] \nonumber\\
  +\frac{4M^2}{p\cdot q}n_\mu n_\nu&\Big[c_1^q\big(G_{3d}+S_{LL}G_{3dLL}\big)+ c_3^q \lambda_h G_{3dL}\Big]. \label{f:W1t4munu}
\end{align}

It is convenient to divide the contributions from quark-gluon-gluon-quark correlator $\hat \Xi^{(2)}_{\rho\sigma}$ into two parts, one is the middle-cut and the other is the left- and right-cut part. We first consider the middle-cut part and use the superscript $M$ to distinguish it from the others. Using
\begin{align}
  & \mathrm{Tr}\big[\hat h^{(2)}_{\mu\nu} \gamma^\alpha \big](g_{\perp\rho\alpha} +i\varepsilon_{\perp\rho\alpha} )
 =-\frac{16}{p^+}\big(c_1^q-c_3^q\big)p_\mu p_\nu ,\label{f:traceh2t4} \\
  & \mathrm{Tr}\big[\hat h^{(2)}_{\mu\nu} \gamma^5 \gamma^\alpha \big](i\varepsilon_{\perp\rho\alpha}+g_{\perp\rho\alpha})
 =-\frac{16}{p^+}\big(c_1^q-c_3^q\big) p_\mu p_\nu ,\label{f:traceh25t4}
\end{align}
and Eqs. (\ref{f:Xi2decomp})-(\ref{f:tXi2decomp}) yields
\begin{align}
  z\tilde W^{(2)M}_{t4\mu\nu} = -\frac{4M^2}{(p\cdot q)^2} p_\mu p_\nu
  &\Big[c_1^q\big(D^M_{3dd}+S_{LL}D^M_{3ddLL}\big)\nonumber\\
  +&c_3^q \lambda_h D^M_{3ddL}\Big] \nonumber\\
  -\frac{4M^2}{(p\cdot q)^2} p_\mu p_\nu&\Big[c_1^q\big(G^M_{3dd}+S_{LL}G^M_{3ddLL}\big)\nonumber\\
  +& c_3^q \lambda_h G^M_{3ddL}\Big]. \label{f:W2t4munuM}
\end{align}
Here we define $\tilde W_{t4\mu\nu}=\tilde W^{(0)}_{t4\mu\nu}+\tilde W^{(1)L}_{t4\mu\nu}+\left(\tilde W^{(1)L}_{t4\nu\mu}\right)^* + \tilde W^{(2)M}_{t4\mu\nu}$. By using Eqs. (\ref{f:t4eomD})-(\ref{f:t4eomGL}), (\ref{f:W0t4munu}), (\ref{f:W1t4munu}) and (\ref{f:W2t4munuM}), we can obtain
\begin{align}
  z\tilde W_{t4\mu\nu} = \frac{4M^2}{(p\cdot q)^2}& \bar q^\mu \bar q^\nu \Big[c_1^q(D_3 + S_{LL}D_{3LL}) +c_3^q \lambda_h G_{3L}\Big]. \label{f:W012t4M}
\end{align}
It it can be shown that $\tilde W_{t4\mu\nu}$ satisfies the current conservation $q^\mu \tilde W_{t4\mu\nu}=q^\nu \tilde W_{t4\mu\nu}=0$.

To obtain the contributions from the left- and right-cut parts, one need to very carefully to calculate the following traces, 
\begin{align}
  \mathrm{Tr}\big[\hat N^{(2)\rho\sigma}_{\mu\nu}\slashed {\bar n} \big]
  =&+\frac{4(p\cdot q)}{p^+}c^q_1\Big[g_{\perp}^{\rho\sigma}g_{\perp\mu\nu}+g_{\perp[\mu}^{\rho} g_{\perp\nu]}^{\sigma}\Big]
 \nonumber\\
 &-\frac{4(p\cdot q)}{p^+}ic^q_3\Big[g_{\perp\mu}^{\rho}\varepsilon_{\perp\nu}^{~~\sigma} -g_{\perp\nu}^{\sigma}\varepsilon_{\perp\mu}^{~~\rho} \Big],\label{f:traceh2Le}\\
  \mathrm{Tr}\big[\hat N^{(2)\rho\sigma}_{\mu\nu}\gamma^5\slashed {\bar n} \big]
  =&- \frac{4(p\cdot q)}{p^+}c^q_3\Big[g_{\perp}^{\rho\sigma}g_{\perp\mu\nu}+g_{\perp[\mu}^{\rho} g_{\perp\nu]}^{\sigma}\Big]\nonumber\\
  &+\frac{4(p\cdot q)}{p^+}ic^q_1\Big[g_{\perp\mu}^{\rho}\varepsilon_{\perp\nu}^{~~\sigma} -g_{\perp\nu}^{\sigma}\varepsilon_{\perp\mu}^{~~\rho} \Big].\label{f:traceh2Lo}
\end{align}
The hadronic tensor is given by
\begin{align}
  z\tilde W^{(2)L}_{t4\mu\nu} = \frac{2M^2}{p\cdot q}
  \Big[&\big(c_1^qg_{\perp\mu\nu}+ic_3^q\varepsilon_{\perp\mu\nu}\big)\big(D_{3dd}+S_{LL}D_{3ddLL}\big)\nonumber\\
  +&\big(c_3^q g_{\perp\mu\nu}+ic_1^q\varepsilon_{\perp\mu\nu}\big)\lambda_h D_{3ddL}\Big] \nonumber\\
  -\frac{2M^2}{p\cdot q} \Big[&\big(c_1^qg_{\perp\mu\nu}+ic_3^q\varepsilon_{\perp\mu\nu}\big)\big(G_{3dd}+S_{LL}G_{3ddLL}\big)\nonumber\\
  +&\big(c_3^q g_{\perp\mu\nu}+ic_1^q\varepsilon_{\perp\mu\nu}\big)\lambda_h G_{3ddL}\Big]. \label{f:W2t4munuL}
\end{align}
We define $\tilde W^{(2)}_{t4\mu\nu}=\tilde W^{(2)L}_{t4\mu\nu}+\left(\tilde W^{(2)L}_{t4\nu\mu}\right)^*$. Using  Eqs. (\ref{f:t4eomD})-(\ref{f:t4eomGL}) to eliminate the non-independent FFs yields
\begin{align}
  z\tilde W^{(2)}_{t4\mu\nu}= \frac{4M^2}{p\cdot q}\Big[&\big(c_1^qg_{\perp\mu\nu}+ic_3^q\varepsilon_{\perp\mu\nu}\big)\big(D_{-3dd}+S_{LL}D_{-3ddLL}\big)\nonumber\\
  +&\big(c_3^q g_{\perp\mu\nu}+ic_1^q\varepsilon_{\perp\mu\nu}\big)\lambda_h D_{-3ddL}\Big], \label{f:W2t4munuLandR}
\end{align}
where
\begin{align}
  &D_{-3dd}=D_{3dd}-G_{3dd}, \\
  &D_{-3ddL}=D_{3ddL}-G_{3ddL}, \\
  &D_{-3ddLL}=D_{3ddLL}-G_{3ddLL}.
\end{align}

In this part we obtain the complete hadronic tensor up to twist-4 level. In Eq. (\ref{f:Wt2munu}) we show the leading twist hadronic tensor while show the twist-3 hadronic tensor in Eq. (\ref{f:Wt3}). The twist-4 hadronic tensors are given in Eqs. (\ref{f:W012t4M}) and (\ref{f:W2t4munuLandR}). All these hadronic tensors satisfy the current conservation law.

\subsection{Contributions from the four-quark correlator}

\begin{figure}[t]
  \centering
  \includegraphics[width=7.5cm]{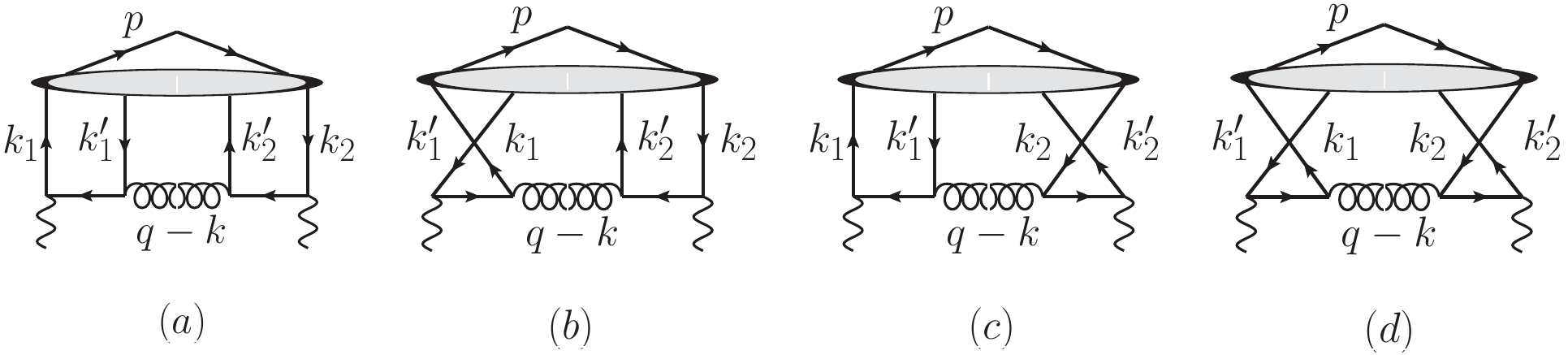}\\
  \caption{The first four of the four-quark diagrams where no multiple gluon scattering is involved.
  In (a), we have $k_1'=k_1-k$ and $k_2'=k_2-k$;
  in (b)  we have the interchange of $k_1$ with $k_1'$;
  in (c) we have the interchange of $k_2$ with $k_2'$;
  in (d) we have both interchanges of $k_1$ with $k_1'$ and $k_2$ with $k_2'$. } \label{SIA4q}
\end{figure}

At twist-4, there are also  contributions from diagrams involving the four-quark correlator \cite{Ellis:1982cd,Ellis:1982wd,Qiu:1988dn} except for those from quark-j-gluon-quark correlators. Following the previous discussion, we consider the four-quark correlator in this part. The general operator definition of the four-quark correlator is given by

\begin{align}
  \hat \Xi^{(0)}_{(4q)}(k_1,k,k_2)&=\frac{g^2}{8}\int\frac{d^4y}{(2\pi)^4}\frac{d^4y_1}{(2\pi)^4}\frac{d^4y_2}{(2\pi)^4}\nonumber\\
  &\times e^{-ik_1y+i(k_1-k)y_1-i(k_2-k)y_2} \nonumber\\
  &\times\sum_X\langle 0|\bar\psi(y_2)\mathcal{L}^\dag(0,y_2)\psi(0)|hX \rangle \nonumber\\
  &\times\langle hX|\bar\psi(y)\mathcal{L}(y,y_1)\psi(y_1)|0 \rangle. \label{f:4qcorr}
\end{align}
Some example of the four-quark diagrams are shown in Fig. \ref{SIA4q}. We note that if the cut is given at the middle we have contributions from the gluon jet. If the cut at the left and/or right, we have contributions from the quark jet. Both of them contribute to the vector meson production annihilation process, in this case we consider them together.

It can be shown that the collinear expansion can also be applied to this case and the gauge links included
in the correlators given by Eq.~(\ref{f:4qcorr}) are obtained by taking the multiple gluon scattering into account. The explicit factorization form of the hadronic tensor is the product of the hard part and the four-quark correlator. It can be written as 
\begin{align}
  \tilde W_{4q\mu\nu}^{(g/q)}=\frac{1}{p\cdot q} &\int dzdz_1dz_2  h^{g/q}_{4q}\Big[\big(c_1^q g_{\perp\mu\nu}+i c_3^q \varepsilon_{\perp\mu\nu} \big)\hat C_s\nonumber\\
  +& \big(c_3^q g_{\perp\mu\nu}+ic_1^q \varepsilon_{\perp\mu\nu}\big)\hat C_{ps}\Big]. \label{f:W4q-uni}
\end{align}
Here we have written the hadronic tensor $W_{4q\mu\nu}^{(g)}$ for both the quark and gluon jet cases into a unified form which is distinguished  by superscripts $g, q$ for gluon and quark jet, respectively.
$\hat C_s$ and $\hat C_{ps}$ are these correlators considered. They can also be written as the unified form
\begin{align}
  &\hat C_j=\int\frac{d^2k'_\perp}{(2\pi)^2}\int d^4k_1d^4kd^4k_2\delta(z-\frac{p^+}{k^+})\delta(k_1^+ z_1-p^+)\nonumber\\
  &\delta(k_2^+ z_2-p^+)(2\pi)^2\delta^2(\vec k_\perp + \vec k_\perp')\hat \Xi^{(0)}_{(4q)j}(k_1,k,k_2;p), \label{f:Cs-ps}
\end{align}
where $j=s, ps$. The corresponding $\hat \Xi^{(0)}_{(4q)s}$ and $\hat \Xi^{(0)}_{(4q)ps}$ are given by
\begin{align}
  \hat \Xi^{(0)}_{(4q)s}&=\frac{g^2}{8}\int\frac{d^4y}{(2\pi)^4}\frac{d^4y_1}{(2\pi)^4}\frac{d^4y_2}{(2\pi)^4} e^{-ik_1y+i(k_1-k)y_1-i(k_2-k)y_2} \nonumber\\
  &\sum_X \Big\{\langle 0|\bar\psi(y_2)\slashed n\psi(0)|hX \rangle \langle hX|\bar\psi(y)\slashed n\psi(y_1)|0 \rangle \nonumber\\
   +&\langle 0|\bar\psi(y_2)\gamma^5\slashed n\psi(0)|hX \rangle \langle hX|\bar\psi(y)\gamma^5\slashed n\psi(y_1)|0 \rangle  \Big\}, \label{f:Xis4q}\\
  \hat \Xi^{(0)}_{(4q)ps}&=\frac{g^2}{8}\int\frac{d^4y}{(2\pi)^4}\frac{d^4y_1}{(2\pi)^4}\frac{d^4y_2}{(2\pi)^4} e^{-ik_1y+i(k_1-k)y_1-i(k_2-k)y_2} \nonumber\\
  &\sum_X \Big\{\langle 0|\bar\psi(y_2)\gamma^5\slashed n\psi(0)|hX \rangle \langle hX|\bar\psi(y)\slashed n\psi(y_1)|0 \rangle \nonumber\\
   +&\langle 0|\bar\psi(y_2)\slashed n\psi(0)|hX \rangle \langle hX|\bar\psi(y)\gamma^5\slashed n\psi(y_1)|0 \rangle  \Big\}. \label{f:Xips4q}
\end{align}
For simplicity, we have omitted gauge links in Eqs. (\ref{f:Xis4q})-(\ref{f:Xips4q}).

In the hadronic tensor shown in Eq. (\ref{f:W4q-uni}), $h^{g/q}_{4q}$ denotes the sum of all the hard parts which correspond to the four diagrams in Fig. \ref{SIA4q}. They are 
\begin{align}
  h_{4q}^{g}&=\frac{z z_B^3 \delta(z-z_B)}{\big(z_1-z_B+i\epsilon\big)\big(z_2-z_B-i\epsilon\big)}+ \frac{z_B^2/z_1 z_2 \delta(z-z_B)}{\big(1/z_1+i\epsilon\big)\big(1/z_2-i\epsilon\big)} \nonumber\\
  &- \frac{z_B^3/z_2\delta(z-z_B)}{(z_1-z_B+i\epsilon)(1/z_2-i\epsilon)}-(1\leftrightarrow 2)^*, \label{f:h4gs}\\
  h_{4q}^{qL}&=\frac{z z_B^3 \delta(z_1-z_B)}{\big(z-z_B-i\epsilon\big)\big(z_2-z_B-i\epsilon\big)}- \Big(\frac{1}{z_2}\to \frac{1}{z}-\frac{1}{z_2}\Big) \nonumber\\
  &- \frac{z z_B^3 \delta(z_1 +z_B-\frac{z_1z_B}{z})}{\big(z-z_B-i\epsilon\big)\big(z_2-z_B-i\epsilon\big)} +\Big(\frac{1}{z_2}\to \frac{1}{z}-\frac{1}{z_2}\Big),\label{f:h4qs}
\end{align}
where $z=z_B=p^+/k^+$, $h_{4q}^{qR}(z_1,z,z_2)=h_{4q}^{qL*}(z_2,z,z_1)$.
The complete one is obtained by summing over all the hard parts, i.e. $h_{4q}=h_{4q}^{qL}+h_{4q}^{qR}+h_{4q}^g$.

Equation (\ref{f:W4q-uni}) shows the explicit factorization form of the hadronic tensor. Apart from the tensors, it is convenient to consider the other terms as a whole. As for the quark-$j$-gluon-quark correlators, we decompose $\hat C_s$ and $\hat C_{ps}$ in terms of the four-quark FFs as follows,
\begin{align}
  z\int dzdz_1dz_2 & h_{4q}\hat C_{s}=M^2\big(D_{4q}+S_{LL}D_{4qLL}\big),\label{f:cs4q}\\
  z\int dzdz_1dz_2 & h_{4q}\hat C_{ps}= M^2 \lambda_hG_{4qL}.\label{f:ps4q}
\end{align}
Substituting Eqs. (\ref{f:cs4q})-(\ref{f:ps4q}) into Eq. (\ref{f:W4q-uni}) yields the hadronic tensor for the four-quark correlator contributions,
\begin{align}
  z\tilde W_{4q\mu\nu}&=\frac{M^2}{p\cdot q}\Big[\big(c^q_1 g_{\perp\mu\nu}+ic^q_3 \varepsilon_{\perp\mu\nu}\big)\big(D_{4q}+S_{LL}D_{4qLL}\big) \nonumber\\
  &+ \big(c^q_3 g_{\perp\mu\nu}+ic^q_1 \varepsilon_{\perp\mu\nu}\big) \lambda_h G_{4qL}\Big].\label{f:Wt44q}
\end{align}
We can see that $\tilde W_{4q\mu\nu}$ takes exactly the same form as the leading twist $\tilde W_{t2\mu\nu}$ given in Eq. (\ref{f:Wt2munu}).

\subsection{The cross section at twist-4}

Contracting the leptonic tensor and the hadronic tensor yields the complete cross section of the  vector meson production inclusive electron positron annihilation process. We show the leading twist, twist-3 and twist-4 contributions together,
\begin{align}
  \frac{d\sigma^{ZZ}}{dzdy}&=\frac{2\pi\alpha_{em}^2}{Q^2} \chi \bigg\{ T^q_{0,ZZ}(y)\big(D_1+S_{LL}D_{1LL}\big) + T^q_{1,ZZ}(y)\lambda_h G_{1L} \nonumber\\
  &-\frac{\kappa_M}{z}|S_T|\big(T^q_{2,ZZ}(y)\sin\varphi_S D_T + T^q_{3,ZZ}(y)\cos\varphi_S G_T\big)\nonumber\\
  &+\frac{\kappa_M}{z}|S_{LT}|\big(T^q_{2,ZZ}(y)\cos\varphi_{LT} D_{LT} + T^q_{3,ZZ}(y)\sin\varphi_{LT} G_{LT}\big)\nonumber\\
  &+\frac{2\kappa_M^2}{z^2}D^2(y) c_1^ec_1^q (D_3+S_{LL}D_{3LL}) \nonumber\\
  &+\frac{2\kappa_M^2}{z^2}D^2(y) c_1^ec_3^q \lambda_h G_{3L} \nonumber\\
  &-\frac{4\kappa_M^2}{z}T^q_{0,ZZ}(y)\mathrm{Re}\big(D_{-3dd}+S_{LL}D_{-3ddLL}\big) \nonumber\\
  &-\frac{4\kappa_M^2}{z}T^q_{1,ZZ}(y)\mathrm{Re}\lambda_h G_{-3ddL} \nonumber\\
  &-\frac{\kappa_M^2}{z}T^q_{0,ZZ}(y)\big(D_{4q}+S_{LL}D_{4qLL}\big) \nonumber\\
  &-\frac{\kappa_M^2}{z}T^q_{1,ZZ}(y)\lambda_h G_{4qL}
  \bigg\}, \label{f:cspartonmodel}
\end{align}
where the four-quark correlator contributions are included. To simplify this expression  we have defined $\kappa_M=M/Q$ and
\begin{align}
  &T^q_{0,ZZ}(y)=c_1^ec_1^qA(y)+c_3^ec_3^q B(y), \label{f:T0zz} \\
  &T^q_{1,ZZ}(y)=c_1^ec_3^qA(y)+c_3^ec_1^q B(y), \label{f:T1zz} \\
  &T^q_{2,ZZ}(y)=c_1^ec_1^qC(y)+c_3^ec_3^q D(y), \label{f:T2zz} \\
  &T^q_{3,ZZ}(y)=c_1^ec_3^qC(y)+c_3^ec_1^q D(y). \label{f:T3zz}
\end{align}
Here we only show the cross section of the weak interaction term. The complete cross section also includes the electromagnetic and interference terms, see Eq. (\ref{f:crosssum}). For the electromagnetic interaction, we require $c_3=0$ and $c_1=1$. In this case, only $T^q_{0,ZZ}(y)$ and $T^q_{2,ZZ}(y)$ are left, $T^q_{0,ZZ}(y)\to A(y)$ and $T^q_{2,ZZ}(y)\to C(y)$. For the interference terms, we need to set $c_3=c_A$ and $c_1=c_V$. To be explicit, we have
\begin{align}
  &T^q_{0,\gamma \gamma}(y)= A(y), \label{f:T0gz} \\
  &T^q_{1,\gamma \gamma}(y)=0, \label{f:T1gz} \\
  &T^q_{2,\gamma \gamma}(y)=C(y), \label{f:T2gz} \\
  &T^q_{3,\gamma \gamma}(y)=0. \label{f:T3gz}
\end{align}
\begin{align}
  &T^q_{0,\gamma Z}(y)=c_V^ec_V^qA(y)+c_A^ec_A^q B(y), \label{f:T0gz} \\
  &T^q_{1,\gamma Z}(y)=c_V^ec_A^qA(y)+c_A^ec_V^q B(y), \label{f:T1gz} \\
  &T^q_{2,\gamma Z}(y)=c_V^ec_V^qC(y)+c_A^ec_A^q D(y), \label{f:T2gz} \\
  &T^q_{3,\gamma Z}(y)=c_V^ec_A^qC(y)+c_A^ec_V^q D(y). \label{f:T3gz}
\end{align}
Correspondingly the kinematic factor $\chi$ should changes to $e_q^2$ and $\chi_{int}$ for electromagnetic and interference contributions, respectively.

\section{The complete results at twist-4} \label{sec:results}

\subsection{The structure functions}

In the following, we present structure functions in terms of  FFs by only considering the weak contribution. Other contributions from electromagnetic and interference terms can be obtained by replacing the corresponding factors.
Among the 19 structure functions shown in Eqs. (\ref{f:FUin})-(\ref{f:tFTTin}), six of them have leading twist contributions, they are given by
\begin{align}
  &zF_{U1}=c^e_1 c^q_1\Big[D_1 -\frac{\kappa_M^2}{z}\big(4\mathrm{Re}  D_{-3dd}+ D_{4q}\big) \Big],\label{f:SFun1}\\
  &zF_{U3}=2c^e_3 c^q_3\Big[D_1 -\frac{\kappa_M^2}{z}\big( 4\mathrm{Re} D_{-3dd}+ D_{4q}\big) \Big],\label{f:SFun3}\\
  &z\tilde F_{L1}=c^e_1 c^q_3 \Big[G_{1L}-\frac{\kappa_M^2}{z}\big(4\mathrm{Re}D_{-3ddL}+ G_{4qL}\big) \Big],\label{f:SFL1}\\
  &z\tilde F_{L3}=2c^e_3 c^q_1\Big[G_{1L}-\frac{\kappa_M^2}{z}\big(4\mathrm{Re}D_{-3ddL}+ G_{4qL}\big)\Big],\label{f:SFL3}\\
  &zF_{LL1}=c^e_1 c^q_1 \Big[D_{1LL} -\frac{\kappa_M^2}{z}\big(4\mathrm{Re}D_{-3ddLL}+ G_{4qLL}\big) \Big],\label{f:SFLL1}\\
  &zF_{LL3}=2c^e_3 c^q_3\Big[ D_{1LL} -\frac{\kappa_M^2}{z}\big(4\mathrm{Re}D_{-3ddLL}+ G_{4qLL}\big) \Big].\label{f:SFLL3}
\end{align}
We see that they are functions for the unpolarized, the longitudinally polarized cases.
In Eqs. (\ref{f:FUin})-(\ref{f:tFTTin}) they just correspond to the $(1+\cos^2\theta)$- and $\cos\theta$-terms. We can see that the four-quark correlator contributions are included in Eqs. (\ref{f:SFun1})-(\ref{f:SFLL3}) and they have the same modes as for the leading twist contributions.

There are eight structure functions which have twist-3 contributions, and are given by
\begin{align}
  &z^2F_{T1}^{\sin\varphi_S}=-2\kappa_Mc^e_3 c^q_3 D_T,\label{f:SFT1}\\
  &z^2F_{T2}^{\sin\varphi_S}=-\kappa_Mc^e_1 c^q_1 D_T,\label{f:SFT2}\\
  &z^2\tilde F_{T1}^{\cos\varphi_S}=-2\kappa_Mc^e_3 c^q_1 G_T,\label{f:SFtT1}\\
  &z^2\tilde F_{T2}^{\cos\varphi_S}=-\kappa_Mc^e_1 c^q_3 G_T,\label{f:SFtT2}\\
  &z^2\tilde F_{LT1}^{\sin\varphi_{LT}} = 2\kappa_Mc^e_3c^q_1 G_{LT},\label{f:SFtLT1}\\
  &z^2\tilde F_{LT2}^{\sin\varphi_{LT}} = \kappa_Mc^e_1c^q_3 G_{LT},\label{f:SFtLT2}\\
  &z^2F_{LT1}^{\cos\varphi_{LT}} = 2\kappa_Mc^e_3c^q_3 D_{LT},\label{f:SFLT1}\\
  &z^2F_{LT2}^{\cos\varphi_{LT}} = \kappa_Mc^e_1c^q_1 D_{LT}. \label{f:SFLT2}
\end{align}
They all correspond to the transverse components of hadron polarizations.
They correspond to the $\sin\theta$- and $\sin2\theta$-terms in Eqs. (\ref{f:FUin})-(\ref{f:tFTTin}).

The left three structure functions have only twist-4 contributions, and they are given by
\begin{align}
  &zF_{U2}=2\kappa_M^2 c^e_1 c^q_1 D_3/z^2,\label{f:SFun2}\\
  &z\tilde F_{L2}=2\kappa_M^2 c^e_1 c^q_3 G_{3L}/z^2,\label{f:SFL2}\\
  &zF_{LL2}=2\kappa_M^2 c^e_1 c^q_1 D_{3LL}/z^2. \label{f:SFLL2}
\end{align}
We note that $F_{TT}^{\cos2\varphi_{TT}}$ and $F_{TT}^{\sin2\varphi_{TT}}$ do not have correspondence to the FFs. These structure functions indicates that transverse momentum dependent FFs do not appear in the inclusive annihilation process.

\subsection{The forward-backward asymmetries}

We have emphasized in the $introduction$, the main focus of this paper is calculating the forward-backward asymmetries for the produced hadron in the inclusive annihilation process. 
The forward-backward asymmetry is introduced to describe the angle distribution of the fermions from $Z^0$ decays as introduced in Sec. \ref{sec:definition}. Here we redefine the asymmetry at the hadonic level to illustrate the angle distribution of the produced hadron in the electron positron annihilation process. Comparing to Eq. (\ref{f:fbdef}), we define the forward-backward asymmetry for a hadron as
\begin{align}
 A_{FB}=\frac{\int_0^{1} [d\sigma] d\cos\theta-\int_{-1}^0 [d\sigma] d\cos\theta }{\int_{-1}^1 [d\sigma]_U d\cos\theta}, \label{f:FBtheta}
\end{align}
where $[d\sigma]=d\sigma/dz d\cos\theta$ while $[d\sigma]_U$ denotes the differential cross section for unpolarized case at leading twist. 
In Eq. (\ref{f:cspartonmodel}), the differential cross section is given in terms of $y$ instead of $\cos\theta$, it is then convenient to rewrite the forward-backward asymmetry $A_{FB}$ in the following form,
\begin{align}
 A_{FB}=\frac{\int_{1/2}^{1} (d\sigma) dy-\int_{0}^{1/2} (d\sigma) dy }{\int_{0}^{1} (d\sigma)_U dy}, \label{f:FBy}
\end{align}
where  $(d\sigma)=d\sigma/dz dy$  and $ (d\sigma)_U$ denotes the unpolarized  differential cross section at leading twist only. 
\begin{align}
 \int^1_0  (d\sigma)_U dy= \frac{4\pi \alpha_{em}^2}{3Q^2}\left(e_q^2+\chi c_1^e c_1^q + \chi_{int} c_V^e c_V^q \right) D_1. \label{f:dsigmaggu}
\end{align}

Using the definition in Eq. (\ref{f:FBy}) and the corresponding differential cross section, we obtain,
\begin{align}
  A_{FB,U}&=\frac{3\left(\chi c_3^e c_3^q +\chi_{int} c_A^e c_A^q\right) \tilde D_1}{4\left(e_q^2+\chi c_1^e c_1^q + \chi_{int} c_V^e c_V^q \right) D_1}, \label{f:AD1}\\
  A_{FB,L}&=\frac{3\left(\chi c_3^e c_1^q +\chi_{int} c_A^e c_V^q\right) \tilde G_{1L}}{4\left(e_q^2+\chi c_1^e c_1^q + \chi_{int} c_V^e c_V^q  \right) D_1}, \label{f:AG1L}\\
  A_{FB,LL}&=\frac{3\left(\chi c_3^e c_3^q +\chi_{int} c_A^e c_A^q\right) \tilde D_{1LL}}{4\left(e_q^2+\chi c_1^e c_1^q + \chi_{int} c_V^e c_V^q  \right) D_1}, \label{f:AD1LL}\\
  A_{FB,T}^x&=-\frac{\left(\chi c_1^e c_3^q +\chi_{int} c_V^e c_A^q\right) G_T}{2\left(e_q^2+\chi c_1^e c_1^q + \chi_{int} c_V^e c_V^q  \right) D_1}, \label{f:AGT}\\
  A_{FB,T}^y&=-\frac{\left(\chi c_1^e c_1^q +\chi_{int} c_V^e c_V^q\right) D_{T}}{2\left(e_q^2+\chi c_1^e c_1^q + \chi_{int} c_V^e c_V^q  \right) D_1}, \label{f:ADT}\\
  A_{FB,LT}^x&=\frac{\left(\chi c_1^e c_1^q +\chi_{int} c_V^e c_V^q\right) D_{LT}}{2\left(e_q^2+\chi c_1^e c_1^q + \chi_{int} c_V^e c_V^q  \right) D_1}, \label{f:ADLT}\\
  A_{FB,LT}^y&=\frac{\left(\chi c_1^e c_3^q +\chi_{int} c_V^e c_A^q\right) G_{LT}}{2\left(e_q^2+\chi c_1^e c_1^q + \chi_{int} c_V^e c_V^q  \right) D_1}, \label{f:AGLT}
\end{align}
where subscript $U, L, LL, T$ and $LT$ denote respectively the polarizations of the produced hadron. $\tilde D_1, \tilde G_{1L}$ and $\tilde D_{1LL}$ denote FFs including twist-4 contributions, e.g $\tilde D_1=D_1 -\frac{\kappa_M^2}{z}\big(4\mathrm{Re}  D_{-3dd}+ D_{4q}\big)$.  We can see that Eq. (\ref{f:AD1}) is similar to Eq. (\ref{f:fbmnu}) except for the FF $D_1 (\tilde D_1)$. The explicit factorized forms shown in Eqs. (\ref{f:AD1})-(\ref{f:AGLT}) provide a direct demonstration of the factorization theorem and/or the parton model. They can be used to test the electroweak and strong interactions simultaneously.  These forward-backward asymmetries can also be expressed with structure functions. We do not show them for simplicity. 

\begin{figure}
   \begin{minipage}[t]{1\linewidth}
  \centering
  \includegraphics[width=7cm]{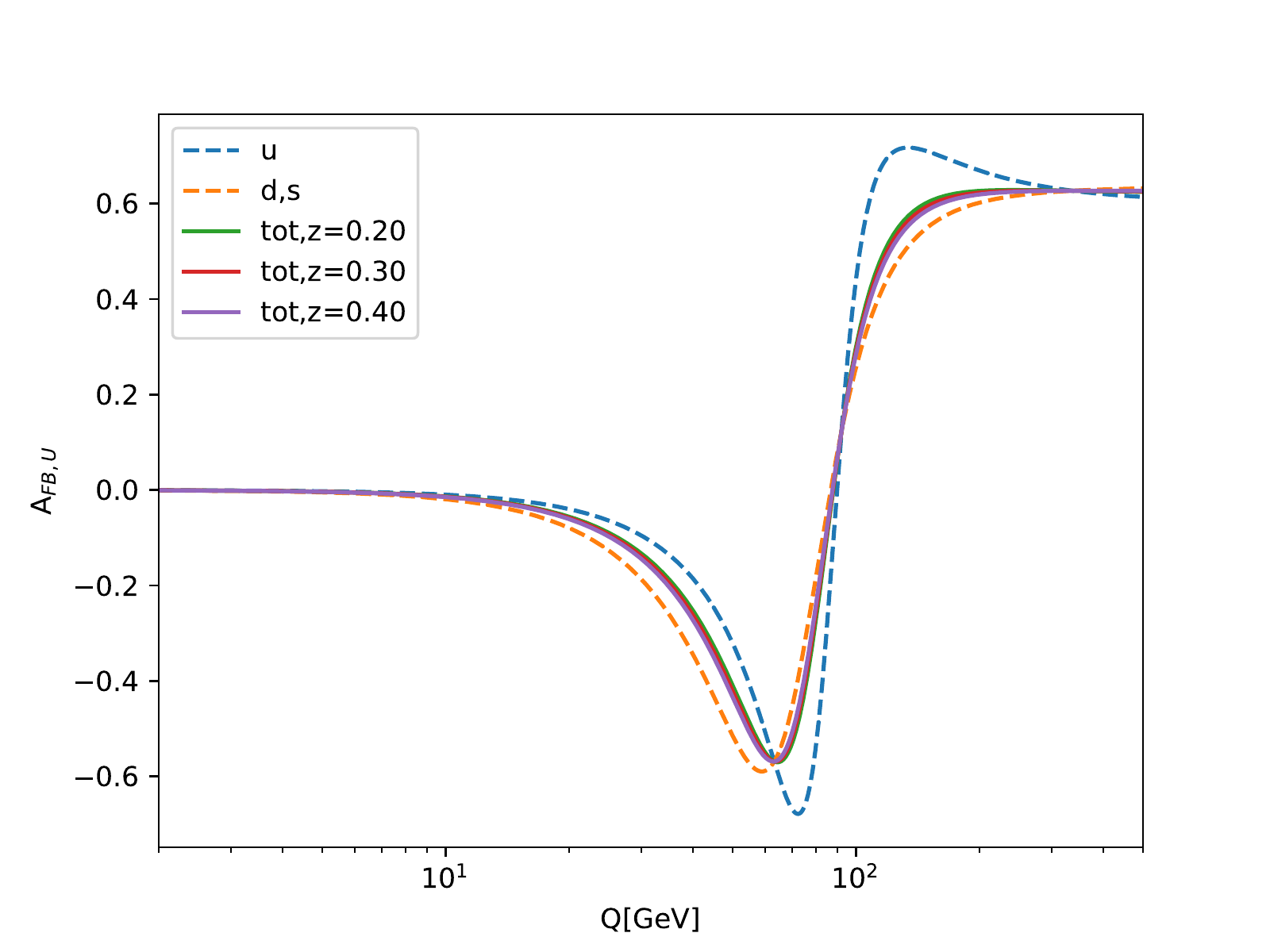}
  \centerline{(a)}
  \end{minipage}
  \begin{minipage}[t]{1\linewidth}
  \centering
  \includegraphics[width=7cm]{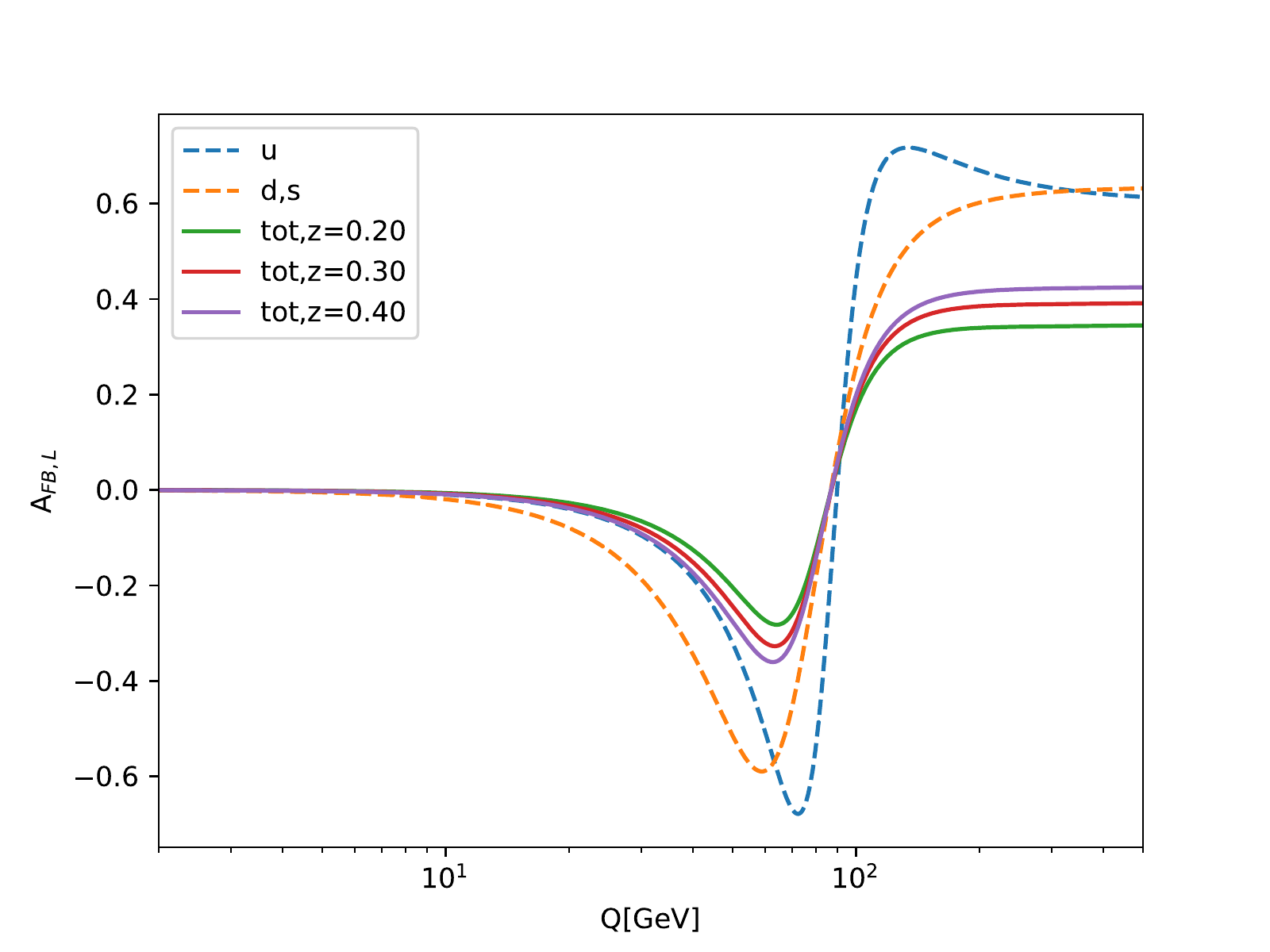}
  \centerline{(b)}
  \end{minipage}
  \caption{The forward-backward asymmetries for $A_{FB,U}$ (a) and   $A_{FB,L}$ (b). Dashed lines show asymmetries for the u and d(s) quarks. Solid lines show asymmetries for produced hadrons with different momentum fraction $z$.} \label{Fab} 
\end{figure}

To have an intuitive impression of the hadron forward-backward asymmetry shown above, we present the numerical values of $ A_{FB,U}$ and $ A_{FB,L}$ in Fig. \ref{Fab}.  The produced hadron is chosen as $\Lambda$ hyperon. Only leading twist contributions are considered. We do not show other asymmetries due to lack of proper parametrizations. The parametrization of the unpolarized FF $D_1$ is taken from AKK08 \cite{Albino:2008fy}. Only the light valence quarks (u, d, s) and gluon are considered here while sea quarks and heavy quarks are ignored (We found that they have limited influences on the numerical results.). The QCD evolution of the FF starts from $Q=2GeV$ and is limited at leading order.  

We use the same parametrization of the longitudinal spin transfer FF $G_{1L}$ given in ref. \cite{Chen:2016iey}. We use 
\begin{align}
 G_{1L}^{s\to \Lambda}(z)=z^a D_{1}^{s\to \Lambda}(z) \label{f:g1ls}
\end{align}
for the $s-$quark FF and 
\begin{align}
 G_{1L}^{q\to \Lambda}(z)=N z^a D_{1}^{q\to \Lambda}(z) \label{f:g1lud}
\end{align}
for the $u-, d-$quark FF, where superscript $q=u, d$.  We fix the parameters as $a=0.5$ and $N= -0.1$. The evolution function and polarized splitting functions can be found in  ref. \cite{Chen:2016iey,Ravindran:1996ri,Ravindran:1996jd}. We do not show them here for simplicity. 

For comparison, we draw asymmetries for quark (u, d, s) as well as that for the produced hadron in the same  panel. We find that they have the same behaviors but different numerical values. This is because the forward-backward asymmetry which arises from the difference of $Z^0$ couplings for left- and right-quarks is dominated by the energy and couplings. At the same time, the longitudinal spin transfer FF $G_{1L}(z)$ satisfies $|G_{1L}(z)|\leq D_1(z)$. The same goes for momentum fractions, $z=0.20, 0.30, 0.40$.  We here only consider the collinear FFs. Parametrizations of the transverse momentum dependent polarizing FF for $\Lambda$ can be found, e.g. in refs. \cite{DAlesio:2020wjq,Callos:2020qtu}.

\subsection{Parity-violating asymmetries}

With the advent of highly polarized electron beams, parity violation measurements have become a standard tool for probing a variety of phenomena. In this part, we calculate the parity-violating asymmetries in the inclusive annihilation process. Parity-violating asymmetry usually describes the difference of the cross section for respectively the right-and left-handed electrons in the deeply inelastic scattering process \cite{Cahn:1977uu,Anselmino:1993tc}. In this paper,  we consider the unpolarized lepton beam and calculate the parity-violating asymmetries with the polarized produced hadron.  The definition of the parity-violating asymmetry is given by 
\begin{align}
 A_{PV}=\frac{ d\sigma (S=+1)-d\sigma (S=-1) }{ d\sigma (S=+1)+d\sigma (S=-1)}, \label{f:PVA}
\end{align}
where $S$ denote the hadon spin, $d\sigma$ denotes the unpolarized differential cross section, i.e. $d\sigma=d\sigma/dzdy$. This definition is different from that in ref. \cite{Chen:2020ugq} where asymmetry was given with respect to the unpolarized electromagnetic cross section. Different definitions in principle do not change the physical meanings. However, numerical result shows the definition  in Eq. (\ref{f:PVA}) is more reasonable. 

\begin{figure}
  \includegraphics[width=7.0cm]{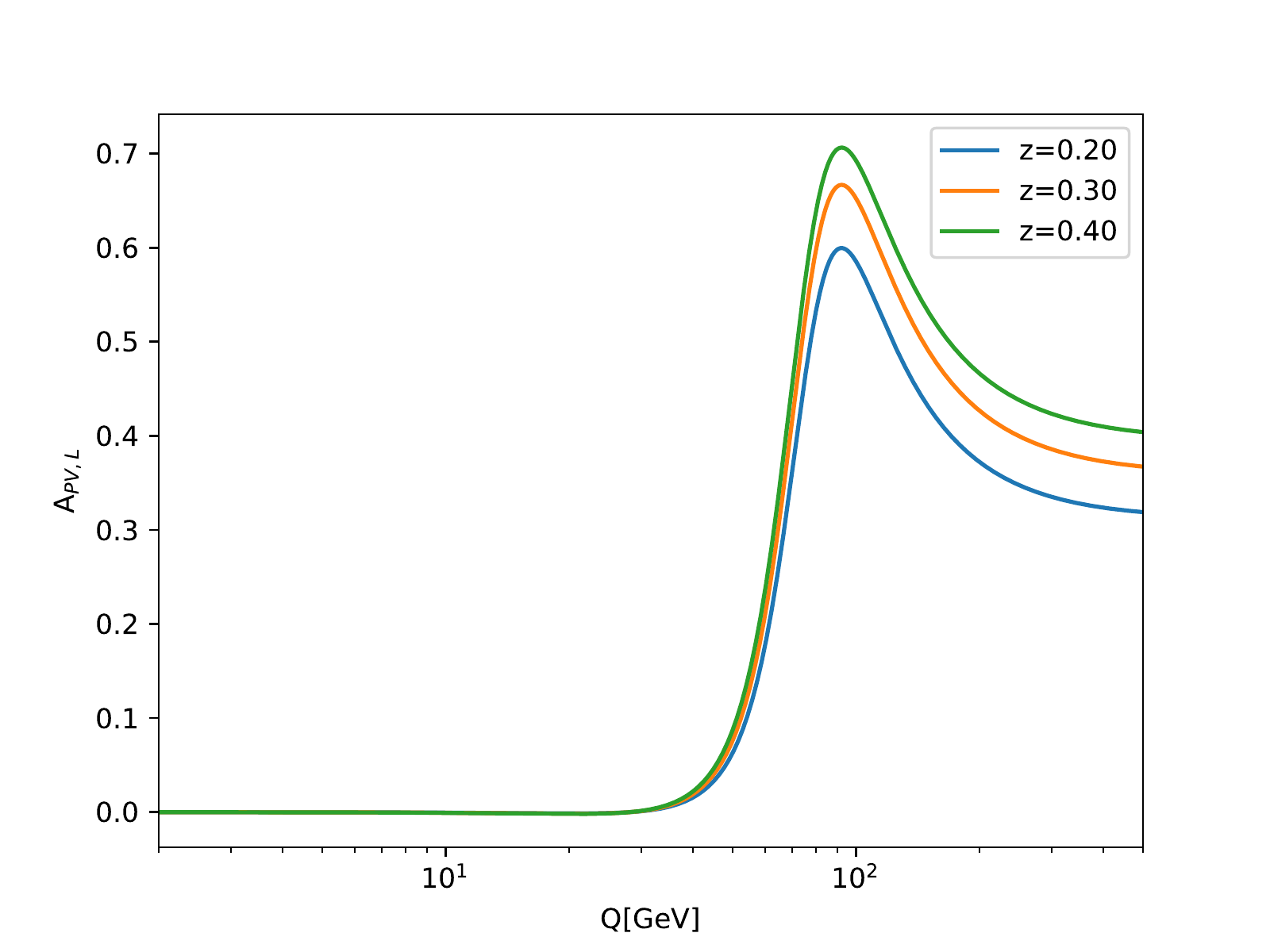}\\
  \caption{The parity-violating asymmetry for $A_{PV,L}$. $z$ is the momentum fraction.}\label{fpv}
\end{figure}

First of all, we present the two asymmetries given by the longitudinal polarized FFs, they are
\begin{align}
 A_{PV,L}&=\frac{\left(\chi T_{1,ZZ}^q+\chi_{int} T_{1,\gamma Z}^q \right) \tilde G_{1L}}{\left(e_q^2 T_{0,\gamma\gamma}^q + \chi T_{0,ZZ}^q + \chi_{int} T_{0,\gamma Z}^q \right)D_1}, \label{f:PVL} \\
 A_{PV,LL}&=\frac{\left(\chi T_{0,ZZ}^q+\chi_{int} T_{0,\gamma Z}^q \right) \tilde D_{1LL}}{\left(e_q^2 T_{0,\gamma\gamma}^q + \chi T_{0,ZZ}^q + \chi_{int} T_{0,\gamma Z}^q \right) D_1}. \label{f:PVLL}
\end{align}
We can see that they are leading twist asymmetries with twist-4 corrections. We use the same parametrization of the longitudinal spin transfer FF $G_{1L}$ shown before and present the numerical values of $ A_{PV,L}$ in Fig. \ref{fpv}.

Correspondingly, there are two twist-4 asymmetries which are given by
\begin{align}
 A^4_{PV,L}&=\kappa_M^2 \frac{2\left(\chi c_1^ e c_3^q+\chi_{int}c_V^ e c_A^q \right) D^2(y) G_{3L}}{z^2\left(e_q^2 T_{0,\gamma\gamma}^q + \chi T_{0,ZZ}^q + \chi_{int} T_{0,\gamma Z}^q \right)D_1}, \label{f:PVLd} \\
 A^4_{PV,LL}&=\kappa_M^2 \frac{2\left(\chi c_1^ e c_1^q+\chi_{int} c_V^ e c_V^q\right) D^2(y) D_{3LL}}{z^2\left(e_q^2 T_{0,\gamma\gamma}^q + \chi T_{0,ZZ}^q + \chi_{int} T_{0,\gamma Z}^q \right)D_1}. \label{f:PVLLd}
\end{align}

We can also calculate the parity-violating asymmetries for the transversely polarized hadron case, they are all twist-3 asymmetries. 
\begin{align}
 A^y_{PV,T}&=-\kappa_M\frac{\left(\chi T_{2,ZZ}^q+\chi_{int} T_{2,\gamma Z}^q \right) D_{T}}{z\left(e_q^2 T_{0,\gamma\gamma}^q + \chi T_{0,ZZ}^q + \chi_{int} T_{0,\gamma Z}^q \right)D_1}, \label{f:PVSy}\\
 A^x_{PV,T}&=-\kappa_M\frac{\left(\chi T_{3,ZZ}^q+\chi_{int} T_{3,\gamma Z}^q \right) G_{T}}{z\left(e_q^2 T_{0,\gamma\gamma}^q + \chi T_{0,ZZ}^q + \chi_{int} T_{0,\gamma Z}^q \right)D_1}, \label{f:PVSx}\\
 A^y_{PV,LT}&=\kappa_M\frac{\left(\chi T_{3,ZZ}^q+\chi_{int} T_{3,\gamma Z}^q \right) G_{LT}}{z\left(e_q^2 T_{0,\gamma\gamma}^q + \chi T_{0,ZZ}^q + \chi_{int} T_{0,\gamma Z}^q \right) D_1}, \label{f:PVLTy}\\
 A^x_{PV,LT}&=\kappa_M\frac{\left(\chi T_{2,ZZ}^q+\chi_{int} T_{2,\gamma Z}^q \right) D_{LT}}{z\left(e_q^2 T_{0,\gamma\gamma}^q + \chi T_{0,ZZ}^q + \chi_{int} T_{0,\gamma Z}^q \right) D_1}. \label{f:PVLTx}
\end{align}
Parity-violating asymmetry which is similar to the forward-backward asymmetry combines the electroweak and QCD theories. Measuring these asymmetries can be important ways to examine electroweak and QCD theories simultaneously.

\section{Summary} \label{sec:summary}

In this paper, we consider the vector meson production in the inclusive electron positron annihilation process and calculate the forward-backward asymmetry in the hadronic level, i.e. the asymmetry in the angular distribution of the produced vector meson. The asymmetry arises from the difference of $Z^0$ couplings for left- and right-handed fermions. Measurements of this asymmetry can enable independent determinations of the neutral-current couplings of these fermions. To deal with the non-perturbative fragmentation process, we present a factorized form of the hadronic tensor by using the collinear expansion method in the parton model.  Results are finally expressed in the factorized forms, see Eqs. (\ref{f:AD1})-(\ref{f:AGLT}).  The explicit factorized forms provide a direct demonstration of the factorization theorem and/or the parton model. We can see that Eq. (\ref{f:AD1}) is similar to Eq. (\ref{f:fbmnu}) except for the FF $D_1 (\tilde D_1)$. 
This process provides not only  a tool for analyzing the hadronic weak interactions but also an opportunity for understanding the parton model of the strong interaction. In other words, the results can be used to test the electroweak and strong interactions simultaneously.  In addition to the forward-backward asymmetries, we also calculate parity-violating asymmetries and structure functions at leading order twist-4.

\section*{Acknowledgements}

{The authors thank Kaibao Chen very much for his kind help. This work was supported by the National Laboratory Foundation (No. 6142004180203).}




 \end{document}